\documentclass[12pt, a4paper]{article}
\usepackage[height=25cm,width=16cm]{geometry}
\usepackage{feynmp}
\usepackage{amsmath}
\usepackage{ascmac}
\usepackage{dcolumn}
\usepackage{bm,here}
\usepackage{subfig}
\usepackage{comment}
\usepackage{ifpdf}
\usepackage{slashed}
\usepackage{colortbl}
\usepackage{color}
\usepackage{ulem}
\usepackage{comment}
\usepackage{braket}
\usepackage[mathscr]{eucal}
\usepackage[sort&compress, numbers, merge]{natbib}

\usepackage{cancel}

\ifpdf
\usepackage{graphicx}     
\usepackage[bookmarksopen,colorlinks=true,linkcolor=bblue,citecolor=ppink,urlcolor=ppink]{hyperref}
\else     
\usepackage[dvipdfmx]{graphicx}     
\usepackage[dvipdfmx,bookmarksopen,colorlinks=true,linkcolor=bblue,citecolor=ppink,urlcolor=ppink]{hyperref}
\fi

\usepackage{multicol}
\definecolor{red}{rgb}{1,0,0}
\definecolor{ppink}{rgb}{0.921545,0.440586,0.687243}
\definecolor{bblue}{rgb}{0.400000,0.400000,1.000000}
\usepackage[charter]{mathdesign}
\usepackage{soul}
\usepackage{wrapfig}

\newcommand\blfootnote[1]{%
	\begingroup
	\renewcommand\thefootnote{}\footnote{#1}%
	\addtocounter{footnote}{-1}%
	\endgroup
}

\begin{document}

\begin{titlepage}

\begin{flushright}
    \hfill LAPTH-029/22
\end{flushright}

\begin{center}

	\vskip 1.5cm
	{\large \bf A Global Analysis of Resonance-enhanced Light Scalar Dark Matter}

	\vskip 2.0cm
	{\large Tobias Binder$^{1,2, *}$\blfootnote{$^*$tobias.binder@tum.de},
			Sreemanti Chakraborti$^{3, \dagger}$\blfootnote{$^\dagger$chakraborti@lapth.cnrs.fr}, \\ [.3em]
			Shigeki Matsumoto$^{1, \ddagger}$\blfootnote{$^\ddagger$shigeki.matsumoto@ipmu.jp}
			and
			Yu Watanabe$^{1, \$}$\blfootnote{$^{\$}$yu.watanabe@ipmu.jp} }
	
	\vskip 2.0cm
	$^1${\sl Kavli IPMU (WPI), UTIAS, U. Tokyo, Kashiwa 277-8583, Chiba, Japan} \\ [.3em]
	$^2${\sl Physik Department T31, James-Franck-Straße 1, Technische Universit\"{a}t M\"{u}nchen, D–85748 Garching, Germany} \\ [.3em]
	$^3${\sl LAPTh, U. Grenoble Alpes, USMB, CNRS, F-74940 Annecy, France} \\ [.3em]
	
    \vskip 3.5cm
    \begin{abstract}
        \noindent
        We study a minimal model for a light scalar dark matter, requiring a light scalar mediator to address the core-cusp problem and interact with the standard model particles. We analyze the model comprehensively by focusing on the Breit-Wigner resonance for dark matter annihilation and self-scattering channels, considering the thermal relic abundance condition that includes the early kinetic decoupling effect, as well as the present and future constraints from collider, direct, and indirect dark matter detections. We found that the scalar dark matter with the mass of 0.3--2\,GeV remains uncharted, which will be efficiently tested by the near future MeV gamma-ray observations.
    \end{abstract}
		
\end{center}
		
\end{titlepage}

\tableofcontents
\newpage
\setcounter{page}{1}
	
\section{Introduction}
\label{sec: intro}

The thermal dark matter, which is often called the weakly interacting massive particle (WIMP), is known to be one of the noteworthy dark matter candidates, because the dark matter density observed today is explained by the so-called freeze-out mechanism\,\cite{Srednicki:1988ce, Bernstein:1985th}, namely the mechanism that explains the history of the Big Bang Nucleosynthesis (BBN) and the recombination in the early universe successfully. Though the dark matter is generally predicted to be in the mass range between ${\cal O}(1)$\,MeV\,\cite{Boehm:2013jpa} and ${\cal O}(100)$\,TeV\,\cite{Griest:1989wd}, that with the mass around the electroweak scale, i.e., 1\,GeV--1\,TeV, have been intensively explored so far because of a possible connection to new physics models for the electroweak naturalness. On the other hand, present negative experimental results (from LHC, direct and indirect dark matter detection experiments) are eroding the parameter space of dark matter with the electroweak mass. It motivates us to consider the thermal dark matter with a lighter ($\lesssim {\cal O}(1)$\,GeV) or heavier ($\gtrsim {\cal O}(1)$\,TeV) mass. We focus on the former case in this article.

In general, light thermal dark matter candidates are severely constrained by the CMB observation. First, as mentioned above, the dark matter candidates are required to be heavier than ${\cal O}(1)$\,MeV not to alter the thermal history of the universe at around the era of the neutrino decoupling\,\cite{Boehm:2013jpa}. Next, the annihilation cross-section of the candidates is also required to be much smaller than 1\,pb at around the era of the recombination not to alter its thermal history\,\cite{Slatyer:2015jla}. However, the annihilation cross-section must be about 1\,pb to satisfy the relic abundance condition via the freeze-out mechanism. So, the candidates are excluded if they are annihilating in the $s$-wave (whenever non-relativistic). Hence, other types of candidates have been proposed to overcome the difficulty, where the dark matter annihilation cross-section is velocity-dependent as it proceeds in the $p$-wave, $s$-channel, etc\,\cite{Lin:2011gj, Diamanti:2013bia, Hara:2021lrj, Bernreuther:2020koj, Kondo:2022lgg,Croon:2020ntf,Wojcik:2021xki}.

In addition to the relic abundance condition and satisfying the CMB constraints, it is of our interest to identify model parameter regions where the dark matter self-scattering cross-section is large enough to alleviate structure formation issues on small scales of the universe. These are, e.g., the core-cusp\,\cite{Moore:1999gc, Donato_2009, de_Blok_2010} and the diversity problem\,\cite{Oman:2015xda}. While combined collisionless dark matter (CDM) and hydrodynamical simulations show that baryonic processes can contribute significantly to core formation (e.g., see ref.\,\cite{Governato_2012}) and therefore could potentially explain the core-cusp issue without the need of dark matter self-interactions, it remains challenging to also explain the experimentally observed diversity of galactic rotation curves with equal maximum rotation speed (diversity problem)\,\cite{Santos_Santos_2020}. It is fair to say that the modeling of baryonic physics is still under intense development and future simulations may shed more light on the baryonic solution of the core-cusp and diversity problem within the CDM paradigm (for more details, see ref.\,\cite{Sales:2022ich}). Here, instead, we consider self-interacting dark matter (SIDM) as an alternative solution of both issues\,\cite{Tulin:2017ara}. In particular, SIDM-only can produce density cores instead of cusps due to the self-interactions leading to thermalization in the inner halo region. When baryons are added in addition, their gravitational potential contributes to the inner thermal shape in the right way to ameliorate the diversity problem\,\cite{Kamada:2016euw,Ren:2018jpt,Creasey_2017, Robertson:2017mgj}. A mild velocity-dependence of the self-scattering cross-section may be needed to be consistent with large-scale structure formation of the universe\,\cite{Peter:2012jh}.

It is also intriguing to see that light dark matter candidates, which predict a velocity-dependent annihilation cross-section, often feature a velocity-dependent self-scattering cross-section which could be sufficiently large to address the core-cusp and the diversity problem on galactic scales while being consistent with CDM large-scale structure formation of the universe. For instance, when the dark matter is a singlet Majorana fermion and couples to a scalar mediator particle, it dominantly annihilates into a pair of mediator particles with a $p$-wave suppressed cross-section. Then, a large and velocity-dependent self-scattering cross-section can also be obtained by exchanging a light mediator particle between the dark matter particles\,\cite{Tulin:2013teo}. Another example is when the dark matter annihilates into standard model (SM) particles via a $s$-channel resonance; its self-scattering cross-section can also be significant and velocity-dependent via the same resonance in the annihilation\,\cite{Duch:2017nbe}.

We consider the latter case in this article. Using a minimal model for a scalar dark matter and a scalar mediator particle, we comprehensively investigate whether it can solve the core-cusp and the diversity problem, consistent with the relic abundance condition and the present and future constraints from collider, direct, and indirect dark matter detections. On the other hand, such a dark matter candidate is known to be ruled out when its mass is around the electroweak scale because the annihilation cross-section at the present universe is also enhanced by the same resonance boosting the self-scattering and, therefore, severely constrained by the indirect dark matter detection\,\cite{Duch:2017nbe}. It is, however, not trivial whether such a candidate is excluded when it is as light as sub-GeV, for, in addition to astrophysical (e.g., the dark matter profile) and theoretical (e.g., the fragmentation function) uncertainties, the typical energy of the signal at the indirect dark matter detection is in the MeV range, which is the one not intensively explored compared to other ranges\,\cite{Greiner:2011ih}.

We first quantitatively figure out the parameter region consistent with the thermal relic abundance condition, the self-scattering condition for the core-cusp and the diversity problem, and the CMB constraints. In calculating the thermal relic abundance of the scalar dark matter, we pay particular attention to the early kinetic decoupling effect\,\cite{Binder:2017rgn} and include it using the DRAKE code\,\cite{Binder:2021bmg}. Then, we consider present and future-expected constraints on the scalar dark matter from collider, direct and indirect dark matter detections, considering several experimental, observational, and theoretical uncertainties to make the analysis robust and conservative. We find that the dark matter with the mass of 0.3--2\,GeV is not excluded yet. It is also found that the surviving region can be efficiently tested by the near future MeV gamma-ray experiments, such as COSI\,\cite{Tomsick:2021wed} and GECCO\,\cite{Orlando:2021get} observations.

This article is organized as follows. In section\,\ref{sec: model}, we show our setup, i.e., the minimal renormalizable model for a light scalar dark matter with a light scalar mediator. We will give all interactions predicted by the model and present formulae for various observables that become phenomenologically important in our analysis. In section\,\ref{sec: Present status}, we quantitatively figure out the model parameter region favored by the relic abundance and the self-scattering conditions, as well as the constraints from the CMB observation. In section\,\ref{sec: detections}, we consider constraints on the scalar dark matter model from collider, direct and indirect detections at present and in the near future. Finally, we summarize our discussion in section\,\ref{sec: conclusion}.

\section{The light scalar dark matter}
\label{sec: model}

The minimal and renormalizable model involving a singlet scalar dark matter, a singlet scalar mediator, and standard model (SM) particles is described by the following Lagrangian: 
\begin{align}
    & {\cal L} = {\cal L}_{\rm SM} + \frac{1}{2} (\partial_\mu \chi)^2 - \frac{\mu_\chi^2}{2} \chi^2
    -\frac{\lambda_{H\chi}}{2} |H|^2 \chi^2 -\frac{\lambda_\chi}{4!} \chi^4
    \nonumber \\
    &\qquad \qquad \qquad \qquad ~~~~ + \frac{1}{2} (\partial_\mu \Phi)^2 - \frac{\mu_{\Phi\chi}}{2} \Phi \chi^2
    - \frac{\lambda_{\Phi \chi}}{4} \Phi^2 \chi^2 - V(\Phi, H), \label{eq: lagrangian} \\
    & V(\Phi, H) = \mu_{\Phi H} \Phi |H|^2 + \frac{\lambda_{\Phi H}}{2} \Phi^2 |H|^2
	+ \mu_1^3 \Phi + \frac{\mu_\Phi^2}{2} \Phi^2 + \frac{\mu_3}{3!} \Phi^3 + \frac{\lambda_\Phi}{4!} \Phi^4,
    \nonumber
\end{align}
where ${\cal L}_{\rm SM}$ is the SM Lagrangian. The fields describing the (real) singlet scalar dark matter and the (real) singlet scalar mediator are denoted by $\chi$ and $\Phi$, respectively, while $H$ is the Higgs doublet. An unbroken discrete $Z_2$ symmetry is imposed in the Lagrangian to make the dark matter stable, where $\chi$ is odd while other fields are even under the symmetry.
 
After the electroweak symmetry breaking and taking the unitary gauge, i.e. $H = (0, v_H + h')^T/\sqrt{2}$ with $v_H \simeq 246$\,GeV being the vacuum expectation value of $H$, while $\Phi = v_\Phi + \phi'$ with $v_\Phi = 0$ being that of $\Phi$,\footnote{We can take the vacuum expectation value of the mediator field to be zero without the loss of generality. Then, the condition $v_\Phi = 0$ gives the relation among the Lagrangian parameters, i.e., $2\mu_1^3 + \mu_{\Phi H} v_H^2 = 0$.} the mediator mixes with the SM Higgs boson. Then, the quadratic terms of $h'$ and $\phi'$ in $V(\Phi, H) + V_H(H)$, with $V_H(H) = (\lambda_H/2)(|H|^2 - v_H^2/2)^2 \subset {\cal L}_{\rm SM}$, are
\begin{align}
    V(\Phi, H) + V_H(H) \, \supset \,
	\frac{1}{2}
	\begin{pmatrix} h' & \phi' \end{pmatrix}
	\begin{pmatrix} m^2_{h' h'} & m^2_{h' \phi'} \\ m^2_{h' \phi'} & m^2_{\phi' \phi'} \end{pmatrix}
	\begin{pmatrix} h' \\ \phi' \end{pmatrix}
	\, = \,
	\frac{1}{2}
	\begin{pmatrix} h & \phi \end{pmatrix}
	\begin{pmatrix} m_h^2 & 0 \\ 0 & m_\phi^2 \end{pmatrix}
	\begin{pmatrix} h \\ \phi \end{pmatrix},
\end{align}
where $m_{h' h'}^2 = \lambda_H v_H^2$, $m_{h' \phi'}^2 = \mu_{\Phi H} v_H$ and $m_{\phi' \phi'}^2 = \mu_\Phi^2 + \lambda_{\Phi H} v_H^2/2$, respectively. The mixing matrix diagonalizing the mass matrix and connecting the states $(h', \phi')$ and $(h, \phi)$ is
\begin{align}
	\begin{pmatrix} h \\ \phi \end{pmatrix} =
	\begin{pmatrix} \cos\theta & -\sin\theta \\ \sin\theta & \cos\theta \end{pmatrix}
	\begin{pmatrix} h' \\ \phi' \end{pmatrix}.
\end{align}
Mass eigenstates and the mixing angle are $m_h^2,\,m_\phi^2 = [(m_{h' h'}^2 + m_{\phi' \phi'}^2) \pm \{(m_{h' h'}^2 - m_{\phi' \phi'}^2)^2 + 4m_{h' \phi'}^2\}^{1/2}]/2$ and $\tan\, (2\theta) = 2m_{h' \phi'}^2/(m_{\phi' \phi'}^2 - m_{h' h'}^2)$ in terms of Lagrangian parameters.

Since the mixing angle is constrained to be small, as we will see in the following sections, $h$ is almost the SM Higgs boson, and its mass is fixed to be $m_h \simeq 125$\,GeV as confirmed at the LHC experiment. Then, all interactions in the model are written as follows:
\begin{align}
    {\cal L}_{\rm int}
	=
	&
	-\frac{C_{h \chi \chi}}{2} h \chi^2
	-\frac{C_{\phi \chi \chi}}{2} \phi \chi^2
	-\frac{C_{h h \chi \chi}}{4} h^2 \chi^2
	-\frac{C_{\phi h \chi \chi}}{2} \phi h \chi^2
	-\frac{C_{\phi \phi \chi \chi}}{4} \phi^2 \chi^2
	-\frac{\lambda_\chi}{4!} \chi^4
	\nonumber \\
	&
	-\frac{s_\theta \phi + c_\theta h}{v_H} \sum_f m_f \bar{f}f
	+\left[\frac{s_\theta \phi + c_\theta h}{v_H} + \frac{(s_\theta \phi + c_\theta h)^2}{2v_H^2}\right] \left(2m_W^2 W_\mu^\dagger W^\mu + m_Z^2 Z_\mu Z^\mu\right)
    \nonumber \\
    &
	-\frac{C_{h h h}}{3!} h^3
	-\frac{C_{\phi h h}}{2} \phi h^2
	-\frac{C_{\phi \phi h}}{2} \phi^2 h
	-\frac{C_{\phi \phi \phi }}{3!} \phi ^3
	\nonumber\\
	&
	-\frac{C_{h h h h}}{4!} h^4
	- \frac{C_{\phi h h h}}{3!} \phi h^3 - \frac{C_{\phi \phi h h}}{4} \phi^2 h^2 -\frac{C_{\phi \phi \phi h}}{3!} \phi^3 h - \frac{C_{\phi \phi \phi \phi}}{4!} \phi^4
	+ \cdots,
	\label{eq: interactions}
\end{align}
where $c_\theta\equiv \cos\theta$, $s_\theta\equiv \sin\theta$ and, ``$\dots$'' represents other interactions in ${\cal L}_{\rm SM}$ which are not affected by the mixing between $H$ and $\Phi$, e.g., gauge interactions of SM fermions, self-interactions of SM gauge bosons, etc. Here, SM fermions (quarks and leptons) are denoted by $f$, while $W_\mu$ and $Z_\mu$ are SM weak gauge bosons ($W$ and $Z$ bosons) with $m_W$ and $m_Z$ being their masses, respectively. Coefficients of dark matter interactions are given as follows:
\begin {align}
    C_{h \chi \chi} &= \lambda_{H \chi} v_H c_\theta - \mu_{\Phi \chi} s_\theta, \nonumber \\
    C_{\phi \chi \chi} &= \lambda_{H \chi} v_H s_\theta + \mu_{\Phi \chi} c_\theta, \nonumber \\
    C_{h h \chi \chi} &= \lambda_{H \chi} c_\theta^2 + \lambda_{\Phi \chi} s_\theta^2, \nonumber \\
    C_{\phi h \chi \chi} &= \lambda_{H \chi} c_\theta s_\theta - \lambda_{\Phi \chi} s_\theta c_\theta, \nonumber \\
    C_{\phi \phi \chi \chi} &= \lambda_{H \chi} s_\theta^2 + \lambda_{\Phi \chi} c_\theta^2.
    \label{eq : couplings}
\end{align}
Other scalar interactions, i.e., those among $h$ and $\phi$ are shown in appendix\,\ref{app: scalar interactions}.

We present below some formulae of physical quantities directly related to our discussion in this paper. In particular, the property of the mediator particle $\phi$, i.e., its partial decay widths, plays the most crucial role. Hence, we summarize those in detail. The annihilation cross-section (into SM particles) and the self-scattering cross-section between the dark matter particles are described by these partial decay widths of the mediator particle.

\subsection{Decay of the mediator particle}
\label{subsec: mediator decay}

The mediator particle behaves like a light SM Higgs boson, as it interacts with SM particles through the mixing between $H$ and $\Phi$. Its partial decay width into SM particles is
\begin{align}
	\Gamma\,(\phi \to {\rm SMs}) =
	\sin^2 \theta\,\left. \Gamma(h_{\rm SM} \to {\rm SMs}) \right|_{m_{h_{\rm SM}}^2 \to m_\phi^2},
	\label{eq: phi decay}
\end{align}
with $h_{\rm SM}$ being the SM Higgs boson. We will discuss each decay mode in more detail.

\noindent
\pmb{$\phi \to \gamma \gamma$:}
The mediator particle decays mainly into two photons when $m_\phi \leq 2m_e$. We use the formula developed in Ref.\,\cite{Leutwyler:1989tn} to calculate its partial decay width $\Gamma(\phi \to \gamma \gamma)$ within the region of $m_\phi \leq 0.6$\,GeV, though this decay channel is not dominant anymore when $m_\phi \geq 2m_e$. The partial decay width is computed using the \texttt{HDECAY} code\,\cite{Djouadi:1997yw} when $m_\phi \geq 2$\,GeV. On the other hand, the formula in Ref.\,\cite{Gunion:1989we} is used in the region of 0.6\,GeV $\leq m_\phi \leq$ 2\,GeV, where it connects the two regions $m_\phi \leq 0.6$\,GeV and $m_\phi \geq 2$\,GeV continuously.

\noindent
\pmb{$\phi \to e^- e^+/\mu^- \mu^+$:}
In the region of $2 m_e \leq m_\phi \leq 2 m_\mu$ with $m_\mu$ being the muon mass, $\phi$ decays mainly into an electron and a positron. Its partial decay width is given by the formula,
\begin{equation}
	\Gamma(\phi \to e^+ e^-) = \sin^2 \theta \times \frac{m_e^2 m_\phi}{8 \pi v_H^2} 
	\left( 1 - \frac{4 m_e^2}{m_\phi^2} \right)^{3/2}.
	\label{eq: electrons}
\end{equation}
When the mediator mass is above the muon threshold but below the pion threshold, namely $2 m_\mu \leq m_\phi \leq 2 m_\pi$, the mediator decays mainly into a pair of muons. Its partial decay width is computed by the same formula as above, with $m_e$ being replaced by $m_\mu$. The formulae for the two channels are used up to $m_\phi = 2$\,GeV. On the other hand, in the mass region of $m_\phi \geq 2$\,GeV, the \texttt{HDECAY} code is used to compute partial decay widths of the channels.

\noindent
\pmb{$\phi \to$ hadrons/($u\bar{u}$, $d\bar{d}$, $s\bar{s}$):}
When the mediator is heavier than twice the pion mass but lighter than a few GeV, it decays mainly into a pair of pions, a pair of $K$ mesons when $m_\phi \geq 2 m_K$, and other hadronic final states, ($4\pi$, $\eta\eta$, $\rho\rho$, etc.) when $m_\phi \gtrsim 1$\,GeV. On the other hand, when the mediator is heavier than a few GeV, the decay of the mediator is well-described by that into a pair of light (i.e., $u$, $d$ and $s$) quarks and a pair of gluons. We adapt the prescription in Ref.\,\cite{Winkler:2018qyg} to calculate the partial decay widths into hadrons and light quarks/gluons, where the mediator is postulated to decay into hadrons when $m_\phi \leq 2$\,GeV, while it decays into light quarks/gluons when $m_\phi \geq 2$\,GeV. The total decay width into all the hadrons is matched with that into light quarks and gluons at $m_\phi = 2$\,GeV by adjusting the normalization of the partial decay width into the hadronic states $4\pi$, $\eta\eta$, $\rho\rho$, etc.\footnote{Partial decay widths of the mediator have been computed at leading order in Ref.\,\cite{Winkler:2018qyg} when $m_\phi \geq 2$\,GeV, while we use the result of the \texttt{HDECAY} code which involves some next leading order corrections to the decay widths. Consequently, our normalization slightly differs from the one in the aforementioned reference.}

\noindent
\pmb{Other decays into SM particles:}
Other decay channels into SM particles open when the mediator becomes heavy; e.g., the decay processes into a pair of tau leptons, charm quarks, bottom quarks, etc. We also take all possible decay channels into account (when the mediator is heavy enough) and compute their partial decay widths using the \texttt{HDECAY} code.

\noindent
\pmb{$\phi \to \chi \chi$:}
In addition to the decay channels into SM particles addressed above, the mediator can decay into a pair of dark matter particles when the mediator mass is larger than twice the dark matter mass, i.e., $m_\phi \geq 2 m_\chi$. Its partial decay width is described by the formula:
\begin{eqnarray}
	\Gamma(\phi \to \chi \chi) = \frac{C_{\phi \chi \chi}^2}{32 \pi m_\phi} \left(1 - \frac{4 m_\chi^2}{m_\phi^2} \right)^{1/2}.
	\label{eq: phi to chi chi}
\end{eqnarray}
The partial decay width is not suppressed by the (small) mixing angle $\sin \theta$, so it often dominates the total decay width of the mediator when this decay channel is available.

\begin{figure}[t]
	\centering
    \includegraphics[keepaspectratio, scale=0.6]{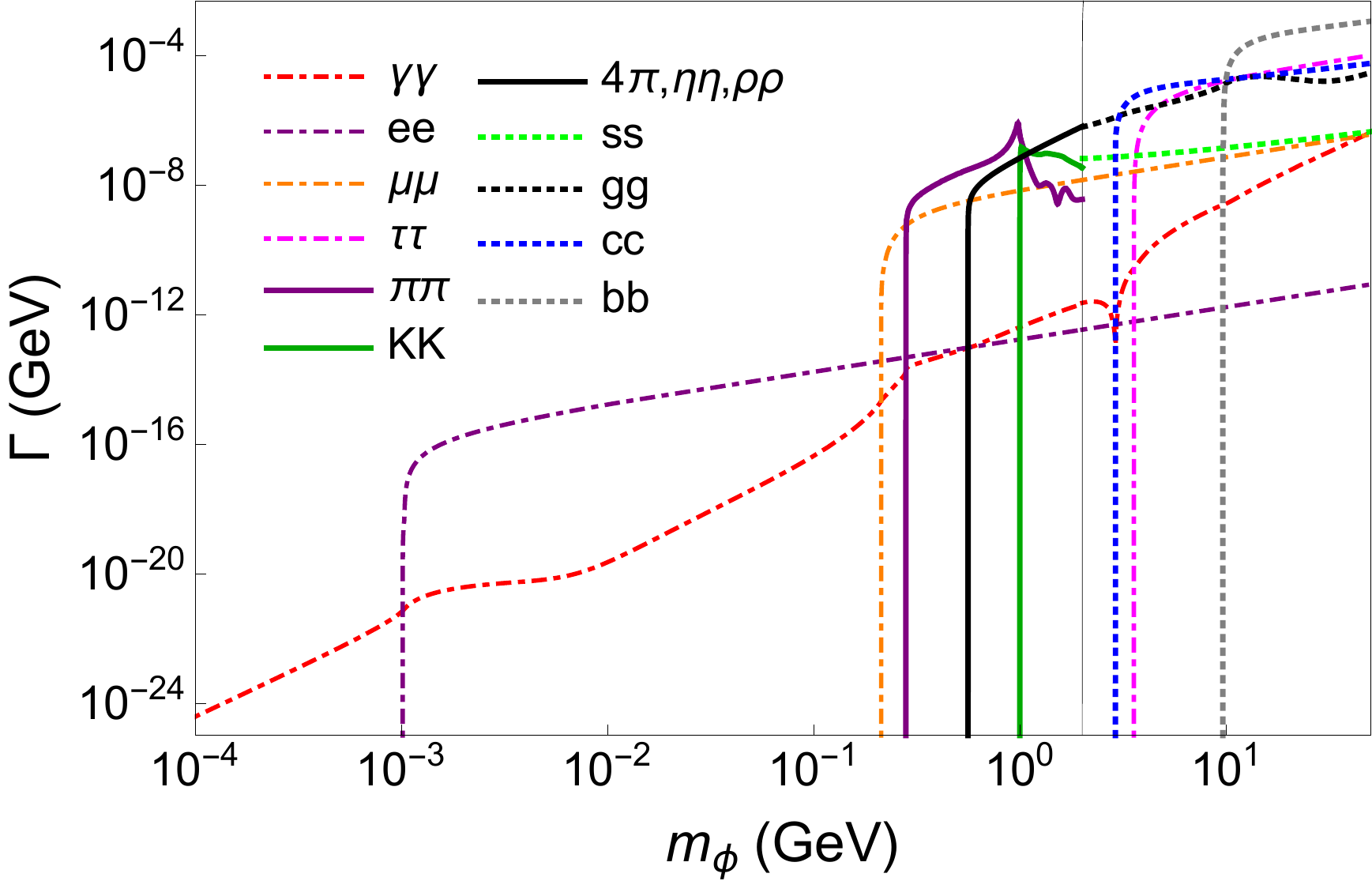}
	\caption{\small \sl Partial widths of the mediator decaying into SM particles assuming $\sin \theta = 1$.}
	\label{fig: phi to SMs}
\end{figure}

All the partial widths of the mediator decaying into SM particles are summarized in Fig.\,\ref{fig: phi to SMs} with the mixing angle being $\sin \theta = 1$ within the mass range of 0.1\,MeV $\leq m_\phi \leq$ 50\,GeV.

\subsection{Dark matter annihilation into SM particles}
\label{subsec: dark matter annihilation}

When we consider the so-called resonance parameter region that we focus on in this paper, i.e., the region satisfying $m_\chi \sim m_\phi/2$, the annihilation cross-section of the dark matter into the SM final state `$f_{\rm SM}$' is described by the corresponding partial decay width of the mediator particle, $\Gamma\,(\phi \to f_{\rm SM})$, and the other partial decay width describing the decay of the mediator particle into a pair of dark matter particles, $\Gamma\,(\phi \to \chi \chi)$, in the following way:
\begin{align}
    &\sigma v\,(\chi \chi \to f_{\rm SM})
    \simeq \frac{2 C_{\phi \chi \chi}^2}{\sqrt{s}} \frac{[\Gamma\,(\phi \to f_{\rm SM})]_{m_\phi^2 \to s}}{(s - m_\phi^2)^2 + s\,\Gamma_\phi^2(s)}
    \simeq \frac{2^{11}\pi}{m_\phi^4 v} \frac{[\Gamma\,(\phi \to f_{\rm SM})\,\Gamma\,(\phi \to \chi \chi)]_{m_\phi^2 \to s}}{(v^2 - v_R^2)^2 + 16\Gamma_\phi^2(s)/m_\phi^2},
    \nonumber \\
    &\Gamma_\phi (s) \equiv [ \Gamma\,(\phi \rightarrow \chi \chi) + \sum_{f_{\rm SM}} \Gamma\,(\phi \to f_{\rm SM}) ]_{m_\phi^2 \to s},
    \label{eq: Ann Xsection}
\end{align}
where $v_R^2 \equiv 4(m_\phi/m_\chi - 2)$, i.e. $m_\chi = (m_\phi/2)/(1 + v_R^2/8)$, and the center of mass energy is $s \simeq m_\chi^2\,(4 + v^2) = m_\phi^2 (1 + v^2/4)/(1 + v_R^2/8)^2$ in the non-relativistic limit of the dark matter.

It is also helpful to discuss the velocity average of the annihilation cross-section (times the relative velocity `$v$') mentioned above, as it is directly related to physical observables such as the thermal relic abundance of the dark matter, various signals at the indirect dark matter detection. The velocity average is estimated in this article using the following formula:
\begin{align}
    \langle \sigma v\,(\chi \chi \to f_{\rm SM}) \rangle_{v_0} &\simeq \int_0^\infty dv\, \sigma v\,(\chi \chi \to f_{\rm SM})\,f(v, v_0),
    \qquad f(v, v_0) \simeq \frac{4 v^2 e^{-v^2/v_0^2}}{\sqrt{\pi}v_0^3},
    \label{eq: velocity average}
\end{align}
where the relative-velocity distribution function of the dark matter $f(v, v_0)$ is normalized to be one, i.e. $\int_0^\infty dv f(v, v_0) = 1$, so that the averaged velocity and the velocity dispersion of the dark matter particle are given by $\langle v \rangle = 2v_0/\sqrt{\pi}$ and $\langle v^2 \rangle = 3v_0^2/2$, respectively.\footnote{Here, we have adopted the Maxwell distribution instead of the Fermi-Dirac/Bose-Einstein distribution for $f(v, v_0)$. This is because the self-scattering condition requires the position of the resonance in the cross-section to be $v_R \sim$ 100\,km/s $\ll c$ as will be discussed in section\,\ref{subsec: SS}, so it does not cause the difference between those using the two distributions. We have numerically confirmed using several parameter sets that the difference is smaller than 1\,\% level, which is much smaller than the uncertainties discussed in sections\,\ref{sec: Present status} and \ref{sec: detections}.}

\subsection{Dark matter self-scattering}
\label{subsec: annihilation cross-section}

Several diagrams contribute to the self-scattering (elastic scattering) between the dark matter particles; the contact diagram (from the dark matter self-interaction proportional to $\lambda_\chi$) and diagrams that $\phi$ or $h$ is exchanged in the $s$-, $t$-, and $u$-channels. The cross-section of the scattering in the non-relativistic limit of the dark matter particles is then estimated as
\begin{align}
    \sigma (\chi \chi \to \chi \chi)
    &= \frac{1}{32 \pi\,s}
    \left[
        \lambda_\chi
        + \frac{C_{\phi \chi \chi}^2}{s - m_\phi^2 + i\sqrt{s}\Gamma_\phi(s)} + \frac{C_{h \chi \chi}^2}{s - m_h^2}
        + \sum_{i = \phi,h} \left( \frac{C_{i \chi \chi}^2}{t - m_i^2} + \frac{C_{i \chi \chi}^2}{u - m_i^2} \right)
    \right]^2,
    \nonumber \\
    &\simeq
    \sigma_0 + \frac{1}{2 \pi m_\phi^6} \frac{C_{\phi \chi \chi}^4}{(v^2 - v_R^2)^2 + 16\Gamma_\phi^2(s)/m_\phi^2}.
\end{align}
Here, we assume $m_\phi \ll m_h$, $v \ll 1$, and drop some interference terms (as it is verified in Ref.\,\cite{Chu:2018fzy}) to obtain the last equation with $\sigma_0 \equiv (\lambda_\chi - 2 C_{\phi \chi \chi}^2/m_\phi^2 - 3C_{h \chi \chi}^2/m_h^2)^2/(32 \pi m_\phi^2)$. Since we will consider the parameter region $m_\chi \simeq m_\phi/2$, we put the imaginary part of the propagator explicitly on the amplitude that the mediator is propagating in the $s$-channel.

As in the case of the annihilation cross-section in subsection\,\ref{subsec: annihilation cross-section}, the velocity averaged cross-section of the self-scattering is directly related to physical observables as
\begin{align}
    \langle \sigma v\,(\chi \chi \to \chi \chi) \rangle_{v_0}
    \simeq \frac{2 v_0}{\sqrt{\pi}}  \sigma_0
    + \frac{1}{2 \pi m_\phi^6} \int^\infty_0 dv \frac{v\,C_{\phi \chi \chi}^4\,f(v, v_0)}{(v^2 - v_R^2)^2 + 16\Gamma_\phi^2(s)/m_\phi^2}.
    \label{eq: self-scattering}
\end{align}
Note that the self-scattering cross-section, $\sigma(\chi \chi \to \chi \chi)$, nearly equals to the one for momentum-transfer, $\sigma_T(\chi \chi \to \chi \chi) \equiv \int d\Omega\,(1 - |\cos\theta|)\,d\sigma(\chi \chi \to \chi \chi)/d\Omega$, for the resonant scattering\,\cite{Chu:2018fzy}, so we do not differentiate between the cross-sections in this article.

\subsection{Dark matter scattering with a nucleon}

The scattering of the dark matter particle ($\chi$) off a nucleon ($N$) occurs by exchanging the mediator particle ($\phi$) or the Higgs boson ($h$) in the $t$-channel. The dominant contribution to the scattering process is a spin-independent one (because the dark matter is a scalar particle), and the explicit form of the cross-section at the zero-momentum transfer limit is
\begin{align}
    \sigma_{\rm SI}(\chi N \to \chi N) =
    \frac{f_N^2\,m_N^4}{4 \pi v_H^2 (m_\chi+m_N)^2}
    \left( \sin \theta \frac{ C_{\phi\chi\chi} }{m_\phi^2} + \cos \theta \frac{ C_{h\chi\chi} }{m_h^2} \right)^2,
    \label{eq: DD}
\end{align}
where $m_N$ is the mass of the nucleon, while $f_N$ is the nuclear from factor that is estimated to be $f_N = 0.284$ and 0.287 when the nucleon $N$ is a neutron and proton, respectively\,\cite{Belanger:2013oya,Thomas:2012tg,Alarcon:2011zs,Alarcon:2012nr}.

\subsection{Invisible Higgs decay width}

The existence of dark matter also contributes to the invisible decay width of the Higgs boson. In the minimal model in eq.\,(\ref{eq: lagrangian}), there are several contributions to the width; one is from the direct decay of the Higgs boson into a dark matter pair, i.e., $h \to \chi \chi$. The next one is the decay of the Higgs boson into a pair of mediator particles that is followed by the invisible decay of the mediator particles, i.e., $h \to \phi \phi \to \chi \chi \chi \chi$. The last one is again from the Higgs decay into a pair of mediator particles, followed by the decay of one of the mediator particles or both mediator particles into SM particles with a long enough decay length. These contributions to the branching ratio of the Higgs invisible decay are
summarized as
\begin{align}
    {\rm Br}\,(h \to \rm inv.)
    &
    = {\rm Br}\,(h \to \chi \chi) 
    + 2 P_{\rm LLP}\, {\rm Br}\,(h \to \phi \phi)\,[1 - {\rm Br}(\phi \to \chi \chi)]\,{\rm Br}\,(\phi \to \chi \chi) \nonumber \\
    &
    + {\rm Br}\,(h \to \phi \phi)\,{\rm Br}\,(\phi \to \chi \chi)^2
    + P_{\rm LLP}^2\,{\rm Br}\,(h \to \phi \phi)\,[1 - {\rm Br}\,(\phi \to \chi \chi)]^2,
    \label{eq: BR_hinv}
\end{align}
where branching ratios of decay modes ${\rm Br}\,(h \to \chi \chi)$ and ${\rm Br}\,(\phi \to \chi \chi)$, and the probability $P_{\rm LLP}$ requiring that the mediator decays outside the detector are given as follows:
\begin{align}
    P_{\rm LLP} &= \exp\,[-\ell_{\rm 
	Detector}/(c \gamma \tau_\phi)], \nonumber \\
	{\rm Br}\,(h \to \chi \chi) &= \frac{1}{\Gamma_h} \Gamma(h \to \chi \chi) = \frac{1}{\Gamma_h} \frac{C_{h \chi \chi}^2}{32 \pi m_h} \left(1 - \frac{4 m_\chi^2}{m_h^2} \right)^{1/2}, \nonumber \\
	{\rm Br}\,(h \to \phi \phi) &= \frac{1}{\Gamma_h} \Gamma(h \to \phi \phi) = \frac{1}{\Gamma_h} \frac{C_{ \phi \phi h}^2}{32 \pi m_h} \left(1 - \frac{4 m_\phi^2}{m_h^2} \right)^{1/2}.
	\label{eq: hinv}
\end{align}
Here, $\Gamma_h = \cos^2\theta\,\Gamma_h^{\rm SM} + \Gamma(h \to \chi \chi) + \Gamma(h \to \phi \phi)$ is the total decay width of the Higgs boson with $\Gamma_h^{\rm SM} \simeq 4.07$\,MeV being the total decay width predicted in the SM. In order to evaluate the probability $P_{\rm LLP}$, we adopt $\ell_{\rm Detector} \simeq 30$\,m as a typical size of the detector at collider experiments, while $c \gamma \tau_\phi \simeq c\,(m_h/2m_\phi)/\Gamma_\phi(m_\phi^2)$ as a typical gamma factor (times the decay length) of the mediator particle produced by the Higgs boson decay at the experiments. We will develop the study of the invisible Higgs decay in more detail in section\,\ref{subsubsec: Higgs production}.

\section{The status of the light scalar dark matter}
\label{sec: Present status}

We figure out the status of the singlet scalar dark matter by showing the parameter region obtained by imposing the self-scattering and relic abundance conditions and constraints from CMB observation. We first summarize model parameters that are convenient to discuss the phenomenology of the scalar dark matter and then discuss the aforementioned conditions and constraints on the minimal model of the dark matter. The constraints and prospects from the direct, indirect, and collider dark matter detection will be discussed later.

\subsection{Model parameters}

It is convenient to use model parameters closely related to physical quantities rather than those used originally in the Lagrangian to define the model. As seen in eq.\,(\ref{eq: lagrangian}), the original model parameters are given by the set, $(\mu_\chi^2, \lambda_{H \chi}, \lambda_\chi, \mu_{\Phi \chi}, \lambda_{\Phi \chi}, \mu_{\Phi H}, \lambda_{\Phi H}, \mu_\Phi^2, \mu_3, \lambda_\Phi)$. Instead, we adopt the following set, $\pmb{(v_R, C_{h \chi \chi}, \sigma_0, \gamma_\phi, C_{\phi \phi \chi \chi}, \sin\theta, C_{\phi \phi h}, m_\phi, C_{\phi \phi \phi}, C_{\phi \phi \phi \phi})}$.

Using the Lagrangian parameters, the dark matter mass is given by $m_\chi^2 = \mu_\chi^2 + \lambda_{H \chi} v_H^2/2$. We, however, use $\pmb{v_R} \equiv 2\,(m_\phi/m_\chi - 2)^{1/2}$ as an input parameter instead of $m_\chi$, because we focus on the resonance parameter region, $m_\chi \sim m_\phi/2$. On the other hand, the coupling between $\chi$ and $\phi$, i.e., $\mu_{\Phi \chi}$, plays the most crucial role in the phenomenology of the light scalar dark matter, and we use the parameter $\pmb{\gamma_\phi} = (C_{\phi \chi \chi}/m_\phi)^2/(64 \pi)$ as an input parameter instead of $\mu_{\Phi \chi}$. Moreover, as seen in section\,\ref{sec: model}, the parameters $\mu_{\Phi \chi}$ and $\mu_\Phi^2$ determine the mass of the mediator particle and the mixing, so we use $\pmb{\sin \theta}$ and $\pmb{m_\phi}$ as input parameters instead of the two parameters. Finally, the couplings $\lambda_{H\chi}$, $\lambda_\chi$, and $\lambda_{\Phi H}$ play dominant roles in the dark matter self-scattering, the direct dark matter detection, and the Higgs decay into a mediator pair, respectively, so we use $\pmb{\sigma_0}$, $\pmb{C_{h \chi \chi}}$, and $\pmb{C_{\phi \phi h}}$ as input parameters instead of those. Physical observables relevant to our discussion depend on these seven parameters.

\subsection{The self-scattering condition}
\label{subsec: SS}

As mentioned in introduction\,(section\,\ref{sec: intro}), the core-cusp and diversity problem is known to be resolved if the dark matter has a large enough self-scattering cross-section, as it thermalizes dark matter particles at the center and, e.g., can make the cuspy structure shallower\,\cite{Spergel:1999mh}.

\begin{figure}[t]
	\centering
    \includegraphics[width=0.6\linewidth]{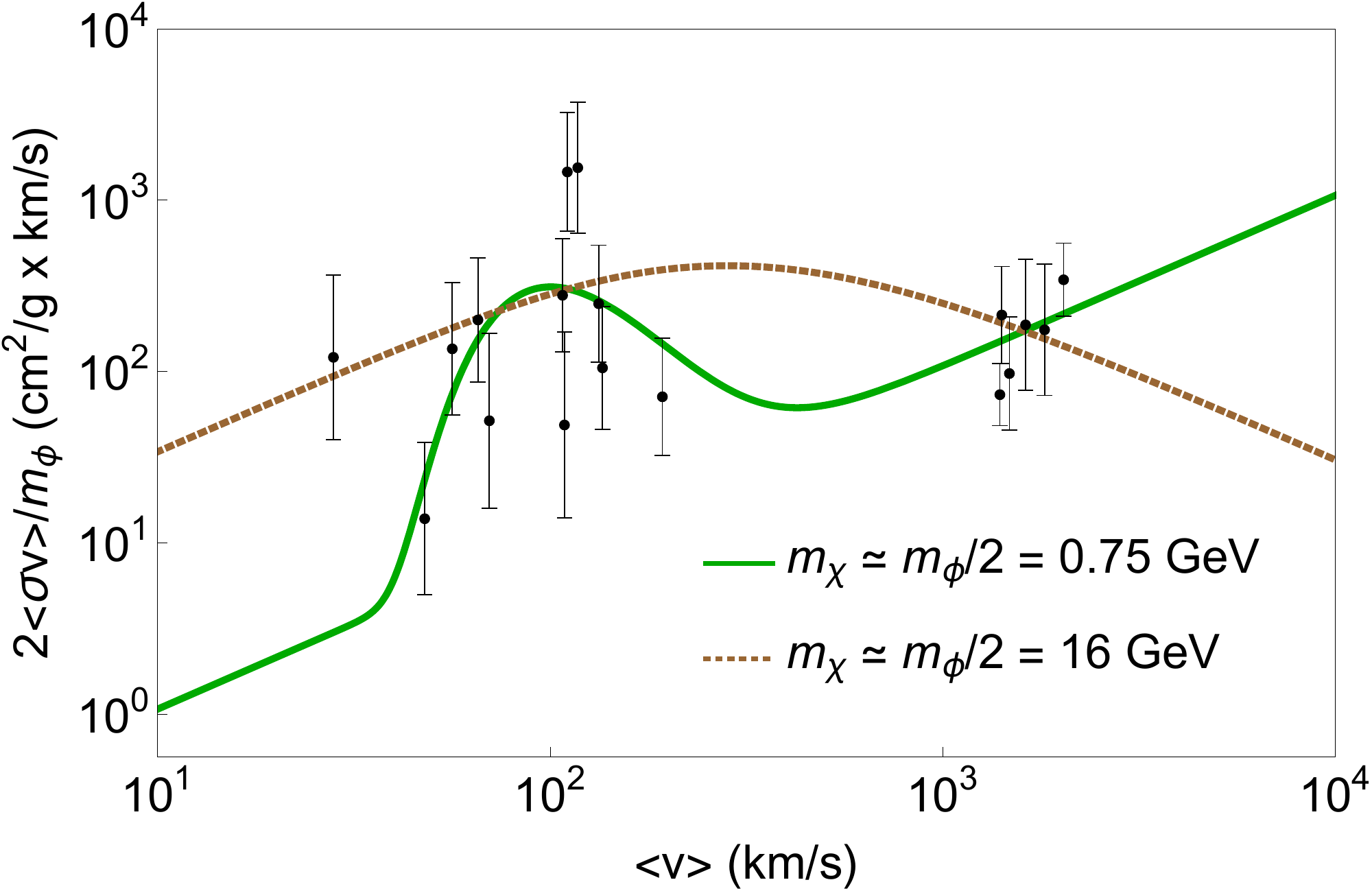}
	\caption{\small \sl The velocity-dependence of the self-scattering cross-section (times the relative velocity) predicted by the scalar dark matter: The prediction using the parameter set, $(v_R, 2\sigma_0/m_\phi, \gamma_\phi, m_\phi) =$ (110\,km/s, 0.1\,cm$^2$/g, 10$^{-9.1}$, 1.5\,GeV), is shown by a green solid line, while that using the set, (5035\,km/s, 0, 10$^{-1.1}$, 32\,GeV), is shown by a brown dashed line. The data obtained by kinematical observation is given by black dots with error bars\,\cite{Kaplinghat:2015aga} . See main text for more detail.}
	\label{fig: self-scattering}
\end{figure}

The strength and the velocity-dependence of the self-scattering cross-section to resolve the problem have been quantitatively discussed in Ref.\,\cite{Kaplinghat:2015aga} as shown in Fig.\,\ref{fig: self-scattering}, where the data is from five clusters\,\cite{Newman:2012nw}, seven low-surface-brightness spiral galaxies\,\cite{KuziodeNaray:2007qi}, and six dwarf galaxies of the THINGS sample\,\cite{Oh:2010ea}. We impose the condition that the model parameter set should give the self-scattering cross-section that is consistent with the data via
\begin{eqnarray}
    -2\ln {\mathcal L}_{\rm SS} = \sum_i
    \left\{
        \frac{(\text{Center value})_i - \log_{10}[2\langle \sigma v\,(\chi \chi \to \chi \chi) \rangle_{v_{0,i}}/m_\phi]}{(\text{Error bar})_i}
    \right\}^2,
\end{eqnarray}
where ${\cal L}_{\rm SS}$ is the part of the likelihood function concerning the self-scattering cross-section involved in our analysis. The index `i' labels a galaxy/cluster involved in the data with a central value and an associated error shown in Fig.\,\ref{fig: self-scattering}, while $\langle \sigma v\,(\chi \chi \to \chi \chi) \rangle_{v_{0,i}}$ is the self-scattering cross-section predicted by the scalar dark matter in eq.\,(\ref{eq: self-scattering}) with $\langle v \rangle_i = 2\,v_{0,i}/\sqrt{\pi}$ being the averaged velocity of the dark matter particle at the "i"-th galaxy/cluster.\footnote{The values of the data should be taken with caution because of the large uncertainties in extracting the cross-sections from the kinematical observation data\,\cite{Sokolenko:2018noz}. Nonetheless, we have adopted those in our analysis to figure out which parameter region does the scalar dark matter prefer to solve the core-cusp problem.}

As shown in Ref.\,\cite{Chu:2018fzy}, the above condition prefers the following two model parameter regions when the total decay width of the mediator particle is governed by the invisible decay into a dark matter pair, i.e., the mixing angle $\sin \theta$ is suppressed:\footnote{Indeed, the mixing angle $\sin\theta$ turns out to be significantly suppressed in the surviving parameter region consistent with existent constraints and the conditions of the self-scattering and the relic abundance.} The first one is the parameter region with the mediator mass of $m_\phi = {\cal O}(10)$\,GeV, where the parameters $v_R$ and $\gamma_\phi$ tend to be required to have a particular correlation to fit the observational data, while the other parameter $\sigma_0$ can be somewhat arbitrary. Indeed, even if we take $\sigma_0 = 0$, the data can be fitted by appropriately choosing $v_R$ and $\gamma_\phi$, as shown by the brown dashed line in Fig.\,\ref{fig: self-scattering}. On the other hand, when the mediator mass is less than about 10\,GeV, the self-scattering condition requires a narrow resonance, i.e., a small $\gamma_\phi$, which should correlate with $m_\phi$ as $\gamma_\phi \propto m_\phi^3$, while the other parameters $v_R$ and $\sigma_0/m_\chi$ are required to be around 100\,km/s, and 0.1\,cm$^2$/g, respectively, as seen by the solid green line in the figure.

\subsection{The relic abundance condition}
\label{subsec: RA}

Next, we consider the condition of the thermal relic abundance assuming that the dark matter abundance observed today is obtained by the freeze-out mechanism of the scalar dark matter in the standard thermal history. This condition can be satisfied in general by choosing an appropriate value of the mixing angle, with the other parameters being those consistent with the previous condition for the dark matter self-scattering discussed in Sec.\,\ref{subsec: SS}. Since the annihilation cross-section of the dark matter is enhanced by the $s$-channel resonance as discussed in Sec.\,\ref{subsec: annihilation cross-section}, the mixing angle $\sin \theta$ is generally required to be very much suppressed to explain the observed dark matter abundance. This reduces the scattering between the dark matter and SM particles at the freeze-out epoch, leading to kinetic decoupling before the chemical one unless other dark matter interactions dominate.\footnote{We consider such a parameter set by considering suppressed $C_{h \chi \chi}$, $C_{\phi \phi h}$ $C_{\phi \phi \chi \chi}$, etc. in this paper.} This early kinetic decoupling is known to alter the calculation of the thermal relic abundance\,\cite{Binder:2017rgn}.

We use the DRAKE code\,\cite{Binder:2021bmg} to calculate the thermal relic abundance of the scalar dark matter, including the effect of the early kinematical decoupling.\footnote{Since the dark matter particles are expected to be self-thermalized by their self-interactions, their distribution function is well-described by the Maxwell-Boltzmann function during the freeze-out epoch.} We also use "the relativistic degrees of freedom" evaluated in Ref.\,\cite{Saikawa:2020swg} instead of that originally implemented in the code, where it has been pointed out in the reference that the uncertainty on the freedom leads to about 10\,\% ambiguity in the relic abundance when the freeze-out temperature is around the QCD phase transition. On the other hand, in order to reduce the computational cost to calculate the relic abundance, we assume that the scattering cross-section between the DM and SM particles is negligibly small during the freeze-out epoch. We have numerically confirmed this fact using several parameter sets. It then turns out that the relic abundance calculated by taking all relevant scattering processes into account is the same as the one computed assuming no scattering between DM and SM particles at around 10\,\% level.\footnote{In the former case including the effect of the scattering, we have calculated the abundance in the largest scattering (i.e., the so-called QCD\,A) scenario suggested in Refs.\,\cite{Gondolo:2012vh, Binder:2017rgn} that makes the effect maximum with the uncertainty caused by the QCD phase transition, i.e., the transition between quark and hadron phases.} We, therefore, adopt the following part of the likelihood function in our analysis,
\begin{eqnarray}
    -2\ln {\mathcal L}_{\rm RA} =
    \left(
        \frac{\Omega_{\rm DM}\,h^2 - \Omega_{\rm TH}\,h^2}{0.2\,\Omega_{\rm DM}\,h^2}
    \right)^2,
\end{eqnarray}
where $\Omega_{\rm DM} h^2 \simeq 0.12$\,\cite{Planck:2018vyg} is the observed dark matter abundance, while $\Omega_{\rm TH}\,h^2$ is the thermal relic abundance predicted by the scalar dark matter. We adopted 20\,\% of $\Omega_{\rm DM} h^2$ as the standard deviation to take the above ambiguities into account conservatively.

One of the consequences obtained by imposing the relic abundance condition (on top of the self-scattering condition) is that the mixing angle $\sin \theta$ is fixed to be an appropriate value. The thermally averaged annihilation cross-section of the scalar dark matter into SM particles in eq.\,(\ref{eq: Ann Xsection}) is given by $\langle \sigma v\,(\chi \chi \to f_{\rm SM}) \rangle \sim \Gamma\,(\phi \to f_{\rm SM}) \Gamma\,(\phi \to \chi \chi)/[\Gamma_\phi(m_\phi)\,m_\phi^3]$ with $\Gamma\,(\phi \to f_{\rm SM}) \propto \sin^2\theta$ in the narrow width approximation, and then the mixing angle is predicted to be about $10^{-3}$, $10^{-5}$, and $10^{-5}$ when $m_\phi \sim$ 100\,MeV, 1\,GeV, and 10\,GeV, respectively. This is because the relic abundance condition requires $\langle \sigma v\,(\chi \chi \to f_{\rm SM}) \rangle \sim$ const., and the invisible decay width of the mediator particle dominates the total decay width $\Gamma_\phi(m_\phi)$ when $m_\phi \gtrsim$ 1\,MeV. Here, the $\sin \theta$ values depend non-trivially on $m_\phi$, as $\Gamma\,(\phi \to f_{\rm SM})$ has a strong $m_\phi$ dependence due to the threshold effect on various decay channels shown in section\,\ref{subsec: mediator decay}.\footnote{The visible decay channels of the mediator particle into SM particles are expected to dominate the total decay width when $m_\phi \ll$ 1\,MeV. This is because the self-scattering condition requires $\Gamma\,(\phi \to \chi \chi) \sim m_\phi \gamma_\phi \propto m_\phi^4$ (as discussed in section\,\ref{subsec: SS}), while the relic abundance condition requires $\Gamma\,(\phi \to f_{\rm SM}) \propto m_\phi^3$ when $m_\phi$ is larger than 1\,MeV. However, the CMB constraint discussed in the next subsection excludes such a parameter region. Hence, the mediator particle always decays invisibly in the parameter region of our interest.} Another consequence is that the predicted thermal relic abundance, including the early kinetic decoupling effect in a $s$-channel resonant region, becomes $\sim 10$ times smaller than that without the effect\,\cite{Binder:2021bmg}, leading to the favored mixing angle evaluated including the effect being $\sim 6$ times smaller than that without it.

\subsection{Constraints from CMB observation}
\label{subsec: CMB}

The introduction of new particles with a mass close to or smaller than the neutrino decoupling temperature $T_D \sim 2$ MeV into the dark sector affects the expansion rate of the universe at the recombination epoch\,\cite{Dolgov:2002wy}. In the scalar dark matter model, the mediator and the dark matter do not interact with neutrinos, so the entropy carried by these particles at the neutrino decoupling temperature is eventually injected only into the electromagnetic plasma at the later universe. Then, it makes the photon temperature larger and the expansion rate of the universe smaller at the recombination epoch than that without such an injection. On the other hand, the CMB observation supports the standard expansion, and it severely constrains the new physics contribution to the expansion rate, leading to the lower limits on the masses of the dark matter and mediator particle not to alter the effective number of relativistic degrees of freedom $N_{\rm eff}$\,\cite{Giovanetti:2021izc, Sabti:2021reh}. Assuming the instantaneous neutrino decoupling and no heating of the neutrinos from electrons and positrons, $N_{\rm eff}$ is given by\,\cite{Matsumoto:2018acr}
\begin{equation}
    N_{\rm eff} \simeq 3 \left\{ 1 + \frac{45}{11\pi^2 T_D^3} \left[ s_\chi(T_D) + s_\phi(T_D) \right] \right\}^{-4/3},
    \qquad
    s_i(T_D) = h_i(T_D) \frac{2\pi^2}{45} T_D^3,
\end{equation}
where $h_i(T_D) = (15 x_i^4)/(4 \pi^4) \int^{\infty}_1 dy \,(4y^2 -1)\sqrt{y^2 -1}/(e^{x_i y}-1)$ with $x_i \equiv m_i/T_D$. Remembering that we consider the resonance parameter region, i.e., $m_\phi \simeq 2 m_\chi$, and taking the latest result of CMB observation on $N_{\rm eff}$, i.e. $N_{\rm eff} = 2.99 \pm 0.17$\,\cite{Planck:2018vyg}, the mass of the mediator particle less than 11\,MeV turns out to be excluded at 95\,\%\,C.L. So, we incorporate the constraint on $m_\phi$ in our analysis using the following part of the likelihood function,
\begin{eqnarray}
    -2 \ln {\cal L}_{N_{\rm eff}} = \left(
        \frac{N_{\rm eff} - 2.99}{0.17}
    \right)^2.
\end{eqnarray}

In addition to the above constraint, the CMB observation is also known to give a constraint on the annihilation cross-section of light thermal dark matter candidates. During the recombination epoch, the annihilation could inject electromagnetically interacting particles (electron, positron, and photon) into the primordial plasma. It increases the residual ionization fraction and modifies the anisotropy of the CMB. On the other hand, the observation prefers the anisotropy without such an injection. Then, the dark matter annihilation cross-section is severely constrained since the CMB anisotropy is precisely observed\,\cite{Planck:2018vyg}. It is also worth emphasizing here that the CMB constraint does not suffer from considerable astrophysical ambiguities, unlike other indirect dark matter detections, since the universe at the recombination epoch is more homogeneous than the present universe, and physics relevant to the constraint is still within the regime of the linear density perturbation.

The CMB constraint on the dark matter annihilation has been discussed in Refs.\,\cite{Slatyer:2015jla, Kawasaki:2021etm}, where an upper limit on $f_{\rm eff\,}(m_\chi)\,\langle \sigma v \rangle_{v_{\rm DM}} /m_\chi$ is given. Here, $\langle \sigma v \rangle_{v_{\rm DM}}$ is the total annihilation cross-section of the dark matter, $\langle \sigma v \rangle_{v_{\rm DM}} \equiv \sum_{f_{\rm SM}} \langle \sigma v\,(\chi \chi \to f_{\rm SM}) \rangle_{v_{\rm DM}}$ with $v_{\rm DM}$ being the typical velocity of the dark matter at the recombination epoch, while $f_{\rm eff\,}(m_\chi)$ is the efficiency of the deposited energy by the annihilation to be injected into the primordial plasma. To a good approximation, the annihilation cross-section is evaluated as $\langle \sigma v \rangle_{v_{\rm DM}} \simeq \sigma v|_{v = 0}$ in the scalar dark matter model because of the following reason: The velocity $v_{\rm DM}$ is estimated to be $v_{\rm DM} \simeq 2 \times 10^{-7}\,( T_\gamma/1\,{\rm eV})\,(1\,{\rm GeV}/m_\chi)\,(10^{-4}/x_{\rm kd})^{1/2}$ with $T_\gamma = 0.235$\,eV and $x_{\rm kd} = T_{\rm kd}/m_\chi$ being the temperature at which the dark matter kinematically decouples from the thermal bath\,\cite{Essig:2013goa}, and $T_{\rm kd}$ is the same order as the freeze-out temperature in the early kinematical decoupling scenario.\footnote{Since the dark matter annihilates into SM particles via the $s$-channel resonance, its annihilation cross-section becomes velocity-dependent at the freeze-out epoch, leading to a slightly larger $v_{\rm DM}$ than that evaluated in this formula. We have numerically confirmed that $v_{\rm DM}$ is yet much smaller than $v_R$ even in the case.} Hence, the typical velocity of the dark matter $v_{\rm DM}$ becomes much smaller than $v_R$ ($\gtrsim$ 100\,km/s) in the parameter region we consider, as discussed in section\,\ref{subsec: SS}, and only the $s$-wave component contributes to the annihilation cross-section.\footnote{A parameter region with $v_R \lesssim v_{\rm DM}$ is also surviving when $m_\phi = {\cal O}(10)$\,GeV, and there the annihilation cross-section $\langle \sigma v \rangle_{v_{\rm DM}}$ becomes larger than $\sigma v|_{v = 0}$. This region is, however, excluded by the CMB constraint discussed in this subsection, even if we adopt the formula $\sigma v|_{v = 0}$ as the annihilation cross-section. So, we apply the "approximation", $\langle \sigma v \rangle_{v_{\rm DM}} \simeq \sigma v|_{v = 0}$, even to such a parameter region, as it is excluded after all.}

\begin{figure}[t]
	\centering
    \includegraphics[width=0.6\linewidth]{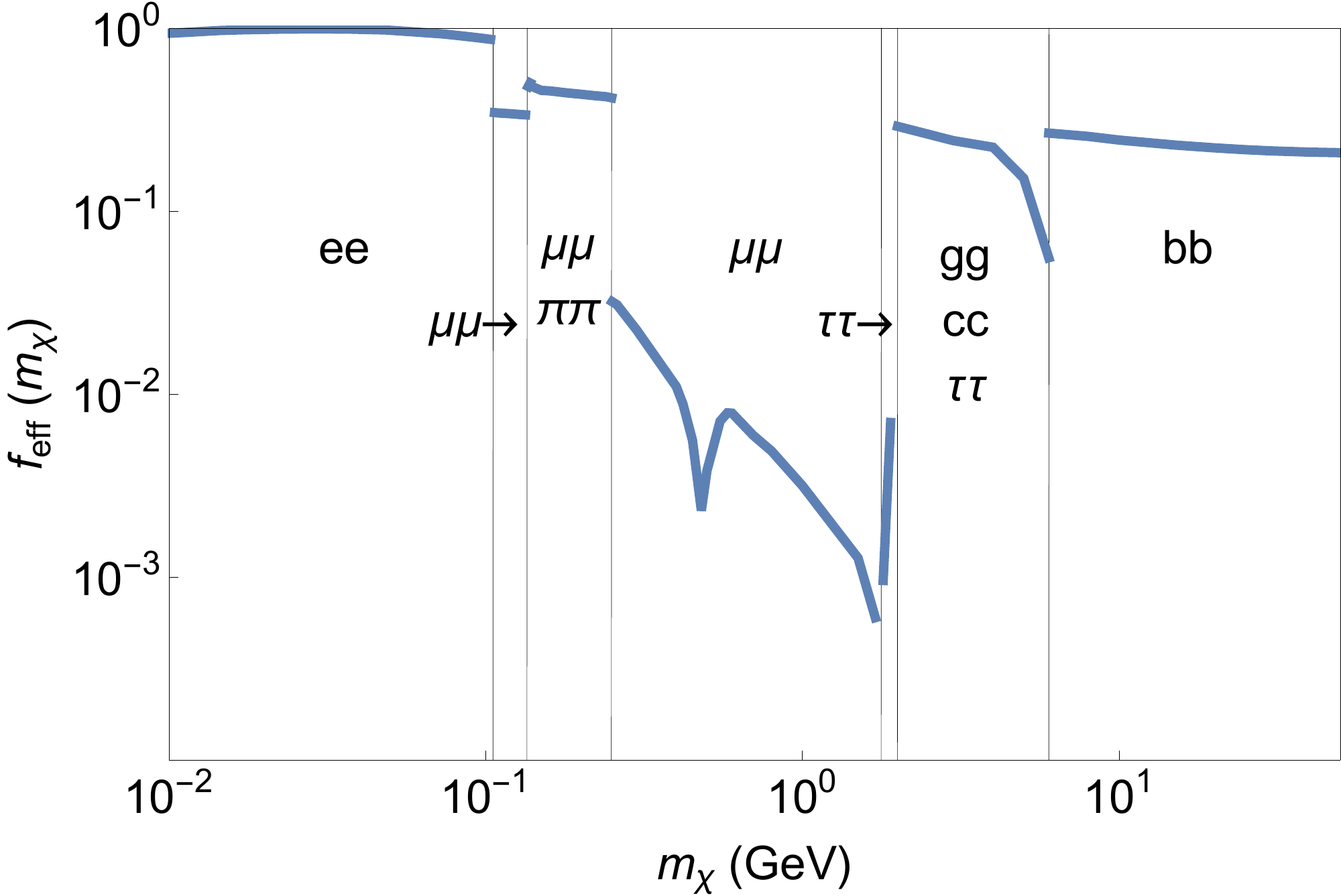}
	\caption{\small \sl The efficiency of the deposited energy from dark matter annihilation to be injected into the primordial plasma is depicted by a solid line. The letters in the figure show the dominant annihilation mode in each mass regions (except the intermediate mass region, 500\,MeV $\leq m_\phi \leq$ 4\,GeV).}
	\label{fig: feff}
\end{figure}

On the other hand, the efficiency $f_{\rm eff\,}(m_\chi)$ is computed by the following formula,
\begin{eqnarray}
    f_{\rm eff\,}(m_\chi) = \int^{m_\chi}_0 dE \frac{E}{2m_\chi} \sum_{f_{\rm SM}} {\rm Br}(\chi \chi \to f_{\rm SM})
    \left[
        2\,f_{\rm eff}^{\,(e)}(E)\,\left. \frac{dN_e}{dE} \right|_{f_{\rm SM}}
        + f_{\rm eff}^{\,(\gamma)}(E)\,\left. \frac{dN_\gamma}{dE} \right|_{f_{\rm SM}}
    \right],
\end{eqnarray}
where ${\rm Br}(\chi \chi \to f_{\rm SM})$ is the branching fraction of the annihilation into the SM final state '$f_{\rm SM}$', while $dN_e/dE|_{f_{\rm SM}}$ and $dN_\gamma/dE|_{f_{\rm SM}}$ are the fragmentation functions describing the number of produced electrons\,(or positrons) and photons with energy E at a given final state '$f_{\rm SM}$'. In addition, $f_{\rm eff}^{\,(e)}(E)$ and $f_{\rm eff}^{\,(\gamma)}(E)$ are the efficiencies to inject the energy into the primordial plasma by a $e^- e^+$ pair and a photon carrying the energies of $2E$ and $E$, respectively\,\cite{Slatyer:2015jla, Kawasaki:2021etm}. In the scalar dark matter model, the efficiency is obtained as shown in Fig.\,\ref{fig: feff}. The letter in the figure shows the dominant annihilation mode of the scalar dark matter in each mass region (except the intermediate mass region, 500\,MeV $\leq m_\phi \leq$ 4\,GeV): The scalar dark matter annihilates into a pair of electrons, muons, and pions/muons when $m_\phi \leq 2m_\mu$, $2 m_\mu \leq m_\phi \leq 2m_\pi$, and $2 m_\pi \leq m_\phi \leq$ 500\,MeV, respectively, where we estimate the efficiency $f_{\rm eff\,}(m_\chi)$, i.e., the fragmentation functions $dN_{e,\,\gamma}/dE|_{f_{\rm SM}}$, using the HAZMA code\,\cite{Coogan:2019qpu}. On the other hand, the dark matter annihilates into a pair of gluons/charm quarks/tau leptons and bottom quarks when 4\,GeV $\leq m_\phi \leq 2m_b$ and $m_\phi \geq 2 m_b$, respectively, with $m_b$ being the bottom quark mass. The efficiency $f_{\rm eff\,}(m_\chi)$ in such mass regions is estimated using the micrOMEGAs code\,\cite{Belanger:2001fz}. In the intermediate-mass region, 500\,MeV $\leq m_\phi \leq$ 4\,GeV, however, there is no robust way to calculate fragmentation functions for hadronic final sates.\footnote{There is a study evaluating the fragmentation functions in this intermediate-mass region for the case of the vector mediator\,\cite{Plehn:2019jeo, Reimitz:2021wcq}, instead of the scalar mediator case that we discuss in this article. Uncertainties associated with the evaluation have also been estimated; those are found to be sizable, as anticipated.} So, as seen in the figure, we estimate the efficiency $f_{\rm eff\,}(m_\chi)$ taking only leptonic final states into account to make our analysis conservative. We will also address how the surviving parameter region changes if we take the efficiency in the intermediate-mass region to be the one obtained by the interpolation from the regions $m_\phi \leq$ 500\,MeV and $m_\phi \geq$ 4\,GeV.

PLANCK collaboration\,\cite{Planck:2015fie} has given an upper limit on the annihilation cross-section as $f_{\rm eff\,}(m_\chi)\,\langle \sigma v \rangle_{v_{\rm DM}}/m_\chi \leq 4.1 \times 10^{-28}$\,cm$^3$/s/GeV at 95\,\%\,C.L. So, we also incorporate this CMB constraint in our analysis using the following part of the likelihood function,
\begin{eqnarray}
    -2 \ln {\cal L}_{\rm CMB} =
    \left[
        \frac{f_{\rm eff\,}(m_\chi)\,\langle \sigma v \rangle_{v_{\rm DM}} /m_\chi}{4.1 \times 10^{-28}/ 1.96}
    \right]^2.
\end{eqnarray}

\subsection{Favored parameter region in the model}
\label{subsec: likelihood result}

\begin{figure}[t]
	\centering
    \includegraphics[width=1\linewidth]{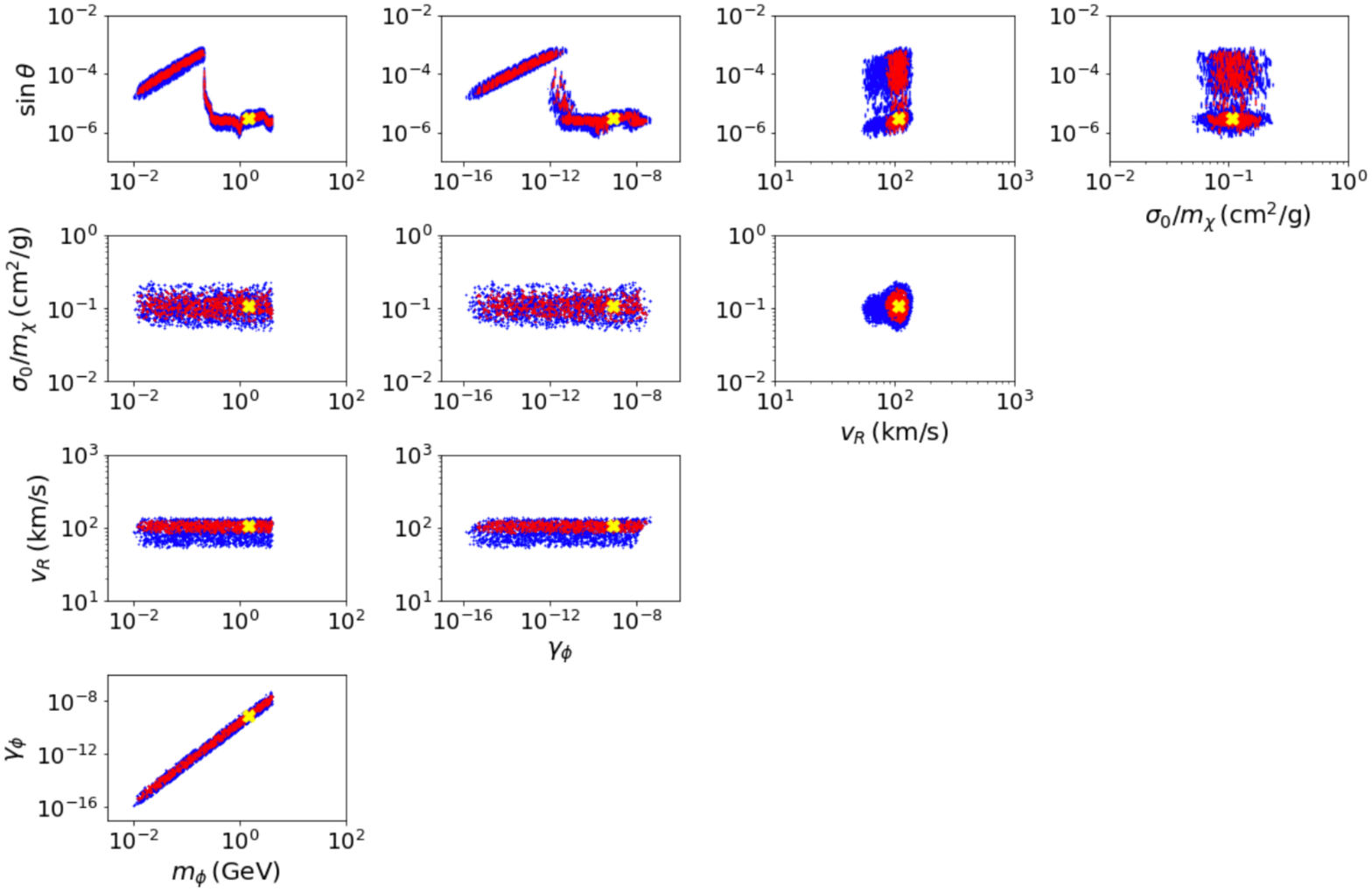}
	\caption{\small \sl The model parameter region favored by the self-scattering condition, the relic abundance condition, and the CMB constraints is shown at 1\,$\sigma$\,(red points) and 2\,$\sigma$\,(blue points) levels with the yellow crossing mark being the parameter set giving the maximum likelihood value.}
	\label{fig: likelihood}
\end{figure}

Here, we present the result of our likelihood analysis in Fig.\,\ref{fig: likelihood}, where the model parameter region favored by the self-scattering condition, the relic abundance condition, and the CMB constraints, is shown at 1\,$\sigma$\,(red points) and 2\,$\sigma$\,(blue points) levels with the yellow crossing mark being the parameter set giving the maximum likelihood value. It can be seen in the figure that the parameters $\sigma_0/m_\chi$ and $v_R$ are required to have ${\cal O}(0.1)$\,cm$^2$/g and ${\cal O}(100)$\,km/s, respectively, while the parameters $m_\phi$ and $\gamma_\phi$ are correlated with each other ($\gamma_\phi \propto m_\phi^3$), as discussed in section\,\ref{subsec: SS}.\footnote{The velocity parameter $v_R$ is almost constant over the entire parameter region in Fig.\,\ref{fig: likelihood} because we need a resonance at around $v_R \simeq$ 100\,km/s for the dark matter self-scattering cross-section, as seen in Fig.\,\ref{fig: self-scattering}.} Moreover, the mixing angle parameter, $\sin \theta$, has a non-trivial correlation with the parameter $m_\phi$ (or $\gamma_\phi$), as discussed in section\,\ref{subsec: RA}; it is $\sin\theta \propto m_\phi$ below the muon threshold ($m_\phi \leq 2m_\mu$) due to the facts $\Gamma_{\rm SM} \propto m_\phi^3$ and eq.\,(\ref{eq: electrons}), while $\sin\theta$ is suppressed above the threshold due to the increase of $\Gamma\,(\phi \to f_{\rm SM})$.

It is also seen in the figure that the mediator mass below $\sim$\,10\,MeV is excluded due to the CMB constraint on $N_{\rm eff}$ discussed in section\,\ref{subsec: CMB}. On the other hand, the mediator mass above $\sim\,4$\,GeV is excluded due to the other CMB constraint on the dark matter annihilation also discussed in section\,\ref{subsec: CMB}. This is because the latter CMB constraint puts an upper limit on $f_{\rm eff\,}(m_\chi)\,\langle \sigma v \rangle_{v_{\rm DM}}/m_\chi$, namely, $2 f_{\rm eff\,}(m_\phi/2)\,\langle \sigma v \rangle_{v_{\rm DM}}/m_\phi$, while the annihilation cross-section behaves as $\langle \sigma v \rangle_{v_{\rm DM}} \sim \Gamma\,(\phi \to \chi \chi)\,\Gamma\,(\phi \to f_{\rm SM})/m_\phi^4 \propto m_\phi^3$, as discussed in Sec.\,\ref{subsec: SS} \& \ref{subsec: RA}, so it eventually leads to an upper limit on the mediator mass.\footnote{$\Gamma(\phi \to \chi\chi) \propto m_\phi^4$ and $\Gamma(\phi \to f_{\rm SM}) \propto m_\phi^3$ are from the self-scattering and the relic abundance conditions.} The aforementioned upper limit $m_\phi \lesssim$ 4\,GeV comes from the fact that we could only take the leptonic final states into account to evaluate the efficiency $f_{\rm eff\,}(m_\chi)$ for the mass region 500\,MeV $\leq m_\phi \leq$ 4\,GeV.\footnote{The upper limit is altered as $m_\phi \lesssim 1$\,GeV if we take the efficiency in the mass region to be the one obtained by the interpolation. Since such a treatment on $f_{\rm eff\,}(m_\chi)$ is hard to be justified at present, the limit mentioned in this footnote should be regarded as one of the referential limits that might be obtained in the future.}

\section{Light scalar dark matter detection prospects}
\label{sec: detections}

This section discusses how the light scalar dark matter predicts strong (or weak) signals at the collider, direct and indirect dark matter detection utilizing the favored model parameter region obtained by the likelihood analysis in the previous section. We discuss both the present status and future expectations of the scalar dark matter probe in each detection.

\subsection{Dark matter detection at colliders}

As seen in the interactions\,(\ref{eq: interactions}), the scalar dark matter model predicts that the dark matter $\chi$ interacts only with the Higgs boson $h$ and the mediator particle $\phi$. On the other hand, the mediator particle interacts with the Higgs boson in various ways and with other SM particles like the light SM Higgs boson with a suppression factor $\sin \theta$. Moreover, the Higgs boson interacts with other SM particles with a small suppression factor $\cos \theta$ compared to the SM case. Hence, we consider collider signals relevant to the Higgs boson production, including the precision Higgs measurement, and the mediator particle production, to discuss the status and prospects of the scalar dark matter scenario for the collider detection.

\subsubsection{Higgs production}
\label{subsubsec: Higgs production}

Compared to the SM case, the production cross-section of the Higgs boson is slightly suppressed by the factor $\cos^2\theta$, and the same factor also suppresses the partial decay widths of the Higgs boson into SM particles. Moreover, the Higgs boson decays into a pair of dark matter and a pair of mediator particles in the scalar dark matter model; those contribute to the invisible decay width of the Higgs boson, as mentioned in section\,\ref{subsec: RA}. As seen in Fig.\,\ref{fig: likelihood}, however, the mixing angle $\sin\theta$ is significantly suppressed, so the deviation from the SM prediction due to the $\cos^2\theta$ factor is negligibly small. Hence, the process to efficiently detect new physics signals in this scenario is from the invisible decay of the Higgs boson. Since no new physics signals are observed so far at the invisible decay, its branching fraction is presently constrained as ${\rm Br}(h \to {\rm inv.}) \leq 0.11$ at 95\,\% C.L.\,\cite{ATLAS:2019cid}, and it will be updated to ${\rm Br}(h \to {\rm inv.}) \leq 0.0026$ if we still do not observe the signals at future lepton colliders such as the international linear collider (ILC)\,\cite{Ishikawa:2019uda}. On the other hand, as seen in eq.\,(\ref{eq: hinv}), the branching fraction of the invisible Higgs boson decay is determined by the couplings $C_{h \chi \chi}$ and $C_{h \phi \phi}$, so it is not directly related to the physics we discussed in the previous section. Namely, the constraint from the invisible decay is always avoided whenever these two couplings are suppressed enough, keeping the result of the previous section intact.

Measuring the aforementioned invisible decay width of the Higgs boson presently constrain the couplings as $(C_{h \chi \chi}^2 + C_{\phi \phi h}^2)^{1/2} \lesssim$ 2.5\,GeV at 95\,\% C.L. when $m_\chi \ll m_h$, and it will be updated to $(C_{h \chi \chi}^2 + C_{\phi \phi h}^2)^{1/2} \lesssim$ 0.37\,GeV if the signal is not observed even in the future. The implication of a non-zero $C_{h \chi \chi}$ coupling is as follows: First, another annihilation channel ($\chi \chi \to h \to$ SMs) starts competing with the resonant channel discussed in the previous section when $C_{h \chi \chi} \geq {\cal O}(1)$\,GeV, because the corresponding cross-section is not suppressed by the small mixing angle $\sin^2 \theta$. Next, we have numerically confirmed that a non-zero $C_{h \chi \chi}$ coupling does not spoil the early kinematic decoupling for the dark matter abundance. Namely, the kinetic decoupling between the dark matter and SM particles still occurs before the chemical one even if we have a non-zero coupling $C_{h \chi \chi} \leq {\cal O}(1)$\,GeV. Instead, we found that the value of the mixing angle squared $\sin^2 \theta$ required by the relic abundance condition could be halved compared to the case with the coupling $C_{h \chi \chi} = 0$ due to the existence of the new annihilation channel. On the other hand, contrary to the case for $C_{h \chi \chi}$, a non-zero $C_{\phi \phi h}$ coupling does not affect the physics discussed in the previous section.

\subsubsection{Mediator production}
\label{subsubsec: Mediator production}

When the mediator particle is lighter than $B$-mesons, it is produced by the decays of the mesons via the sub-process $b \to s \phi$\,\cite{Krnjaic:2015mbs}. Among various $B$-meson decays, one of the efficient channels to search for the mediator particle at $B$-factories is the decay of the charged one, $B^\pm \to K^\pm \phi$. Its partial decay width is estimated by the following formula\,\cite{Batell:2009jf,Dolan:2014ska}:
\begin{equation}
    \Gamma(B^\pm \to K^\pm \phi) = 
    \frac{|C_{sb}|^2 F_K^2 (m_\phi)}{64 \pi m_B^3}
    \left( \frac{m_B^2 - m_K^2}{m_b - m_s} \right)^2
    \sqrt{(m_B^2 - m_K^2 - m_\phi^2)^2 - 4 m_K^2 m_\phi^2},
    \label{eq: BtoKphi}
\end{equation}
where $m_B \simeq$ 5.3\,GeV and $m_K \simeq$ 494\,MeV are $B$- and $K$-meson masses, while $m_s$ is the strange quark mass. The form factor of the $K$-meson is denoted by $F_K(q) \simeq 0.33\,(1 - q^2/38\,{\rm GeV}^2)^{-1}$. On the other hand, the coefficient of the flavor-changing neutral current (FCNC) interaction, $C_{sb}\,\bar{s}_L b_R \phi + {\rm h.c.}$, is obtained as $|C_{sb}| \simeq |3 g^2 m_b m_t^2 V_{ts}^* V_{tb} \sin \theta|/(64 \pi^2 v_H m_W^2) \simeq 6.4 \times 10^{-6}\,|\sin\theta|$ with $V_{tb}$ and $V_{ts}$ being the CKM matrix elements. Combined with the total decay width of the $B$ meson, $\Gamma(B^\pm) \simeq 4.1\times 10^{-13}$ GeV, the above formula allows us to compute the branching fraction of the decay. Since the mediator particle decays invisibly in the favored model parameter region, this charged $B$-meson decay is observed as the one associated with missing energy, $B^\pm \to K^\pm +$ missing, where the background is from the SM process, $B^\pm \to K^\pm + \nu \bar{\nu}$. The Belle\,\cite{Belle:2007vmd} and BaBar\,\cite{BaBar:2010oqg, BaBar:2013npw} collaborations presently put a constraint as ${\rm Br}(B^\pm \to K^\pm \phi) \leq 1.6 \times 10^{-5}$ at 90\,\% C.L., and the Belle\,II collaboration will update the limit as ${\rm Br}(B^\pm \to K^\pm \phi) \leq 5 \times 10^{-7}$ if no signal is observed in the future\,\cite{Manoni:2017lxj}.

When the mediator particle is lighter than $K$-mesons, it is also produced by the decays of the mesons via the other FCNC process $s \to d \phi$\,\cite{Krnjaic:2015mbs}. Among various $K$-meson decays, the decay channels that enable us to efficiently search for the mediator particle are $K^\pm \to \pi^\pm \phi$ and $K_L \to \pi^0 \phi$. Their decay widths are estimated by the following formulae:
\begin{align}
    \Gamma(K^\pm \to \pi^\pm \phi) &=
    \frac{|C_{sd}|^2}{64 \pi m_{K^\pm}^3}
    \left( \frac{m_{K^\pm}^2 - m_{\pi^\pm}^2}{m_s - m_d} \right)^2
    \sqrt{(m_{K^\pm}^2 - m_{\pi^\pm}^2 - m_\phi^2)^2 - 4 m_{\pi^\pm}^2 m_\phi^2},
    \label{eq: KtoPIphi_C} \\
    \Gamma(K_L \to \pi^0 \phi) &=
    \frac{ |C_{sd}|^2}{64 \pi m_{K_L}^3}
    \left( \frac{m_{K_L}^2 - m_{\pi^0}^2}{m_s - m_d} \right)^2
    \sqrt{(m_{K_L}^2 - m_{\pi^0}^2 - m_\phi^2)^2 - 4m_{\pi^0}^2 m_\phi^2}
    \label{eq: KtoPiphi_N},
\end{align}
where $m_{K^\pm} \simeq 494$\,MeV and $m_{K_L} \simeq 498$\,MeV are charged and neutral $K$ meson masses, respectively, while $m_d$ is the down quark mass. Because the scalar form factors of the pions are close to unity\,\cite{Marciano:1996wy}, we do not involve those in the above formulae. The coefficient of the FCNC interaction, $C_{sd}\,\bar{s}_R d_L \phi + {\rm h.c.}$, is given by $|C_{sd}| \simeq |3 g^2 m_s m_t^2 V_{td}^* V_{ts} \sin \theta|/(64 \pi^2 v_H m_W^2) \simeq 1.3 \times 10^{-9}\,|\sin\theta|$ with $V_{ts}$ and $V_{td}$ being the CKM matrix elements. As in the case of the $B$-meson decay, combining with the total decay widths of the $K$-mesons, $\Gamma(K^\pm) \simeq 5.3 \times 10^{-17}$\,GeV and $\Gamma(K_L) \simeq 1.286 \times 10^{-17}$\,GeV, the above formulae enable us to compute the branching fractions of the decays. At present, the E949\,\cite{BNL-E949:2009dza} and NA62\,\cite{NA62:2021zjw} collaborations put a constraint on the branching fraction of the process $K^\pm \to \pi^\pm \phi \to \pi^\pm$ + missing, and the KOTO collaboration\,\cite{KOTO:2018dsc} puts a constraint on the branching fraction of the other process $K_L \to \pi^0 \phi \to \pi^0 +$ missing. In the future, the KLEVER collaboration\,\cite{Beacham:2019nyx} will update the limit on the decay process as ${\rm Br}(K_L \to \pi^0 \phi) \leq 8 \times 10^{-12}$ if no signal is observed there.

\begin{figure}[t]
	\centering
    \includegraphics[width=0.6\linewidth]{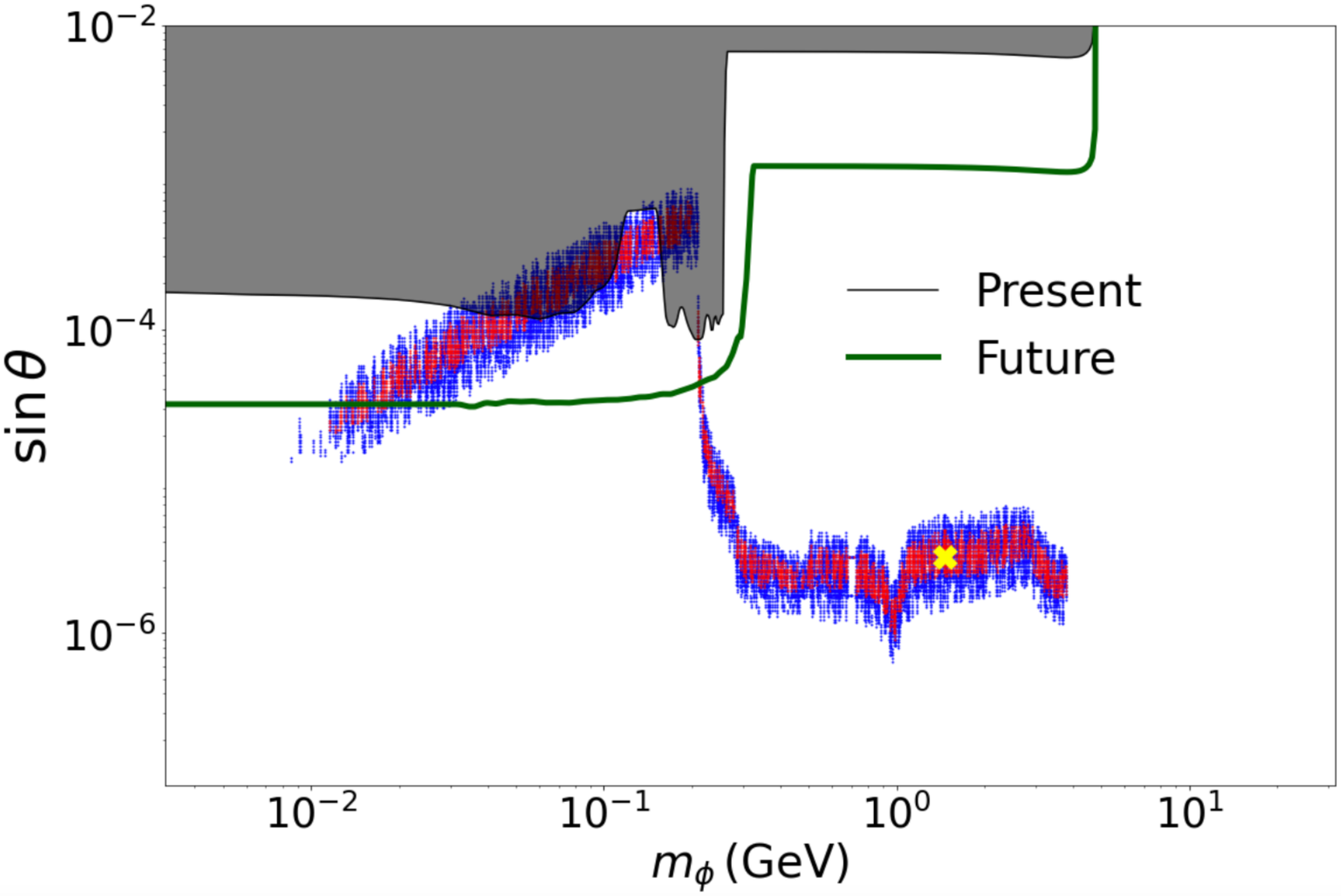}
	\caption{\small \sl The prediction of the scalar dark matter model that is obtained by casting the favored parameter sets in section\,\ref{subsec: likelihood result} onto the $(m_\phi, \sin\theta)$-plane with $C_{h \chi \chi} = C_{\phi \phi h} =$ 0. The colors used for the prediction, i.e., red, blue points, and the yellow crossing mark, are the same as those in Fig.\,\ref{fig: likelihood}. The constraint from present direct dark matter detection experiments and expected sensitivities from future experiments are shown as a gray-shaded region and a dark-green line, respectively.}
	\label{fig: COL}
\end{figure}

As seen in eqs.\,(\ref{eq: BtoKphi})--(\ref{eq: KtoPiphi_N}), the branching fractions of the $B$- and $K$-meson decays depend only on the parameters $m_\phi$ and $\sin \theta$. So, the prediction of the scalar dark matter model on the $(m_\phi, \sin \theta)$-plane is given in Fig.\,\ref{fig: COL}, where it is obtained by casting the favored parameter sets in the previous section onto the plane. The colors used for the prediction, i.e., red, blue points, and the yellow crossing mark, are the same as those used in Fig.\,\ref{fig: likelihood}. The present constraint from the mediator detection experiments (Belle, BaBar, E949, NA62, and KOTO) at 90\,\% C.L. and the expected sensitivities from future experiments (Belle\,II and KLEVER) are shown in the same figure as a gray-shaded region and a dark-green line, respectively\,\cite{2104.07634}: When the mediator particle is lighter enough $K$ meson, i.e., $m_\phi < m_K$, the mixing angle $\sin \theta$ is constrained to be below $\sim 10^{-4}$ mainly by the NA62 experiment.\footnote{The strongest constraint on $\sin \theta$ at the region around $m_\phi \sim 0.1$\,GeV comes from the KOTO experiment.} On the other hand, when $m_K < m_\phi < m_B$, $\sin \theta$ is required to be less than $\sim 10^{-2}$ by both the Belle and BaBar experiments. Future KLEVER and Belle\,II experiments will cover $\sin \theta$ which is ten times better than the present constraints at the regions $m_\phi < m_K$ and $m_K < m_\phi < m_B$, respectively. It is then seen in the figure that some parts of the prediction (parameter region) at $m_\phi \leq 2 m_\mu$ have already been excluded and will be broadly covered in the future.

Non-zero $C_{h \chi \chi}$ and $C_{\phi \phi h}$ couplings do not contribute to the branching fractions of the meson decays that are associated with missing energy, as those give three-body decays, i.e., $B \to K h^* \to K \chi \chi/K \phi \phi$ and $K \to \pi h^* \to \pi \chi \chi/\pi \phi \phi$. In fact, it turns out that the corresponding amplitudes of the three-body decays are negligently small compared to those of the two-body decays, i.e., $B \to K \phi$ and $K \to \pi \phi$, even if the couplings are ${\cal O}(1)$\,GeV. On the other hand, as discussed at the end of section\,\ref{subsubsec: Higgs production}, the mixing angle squared, $\sin^2\theta$, could be halved if $C_{h \chi \chi} = {\cal O}(1)$\,GeV due to the relic abundance condition. The constraint from the meson-decays is, hence, slightly weaker than that in Fig.\,\ref{fig: COL} in such a case.

\subsection{Direct detection of dark matter}
\label{subsec: DD}

Direct dark matter detection is an efficient search strategy for dark matter candidates heavier than sub-GeV. In the scalar dark matter model, the spin-independent (SI) scattering cross-section between the dark matter and a nucleon is given by the formula in eq.\,(\ref{eq: DD}). As seen in the formula, the scattering occurs by exchanging the mediator particle and the Higgs boson in the $t$-channel, so it depends on the coupling constant of the interaction among the mediator or Higgs and dark matter particles, given by $C_{\phi \chi \chi}$ and $C_{h \chi \chi}$, in addition to the other parameters ($m_\phi$, $\sin \theta$, etc.) relevant to the physics discussed in the previous section.

\begin{figure}[t]
	\centering
    \includegraphics[width=0.6\linewidth]{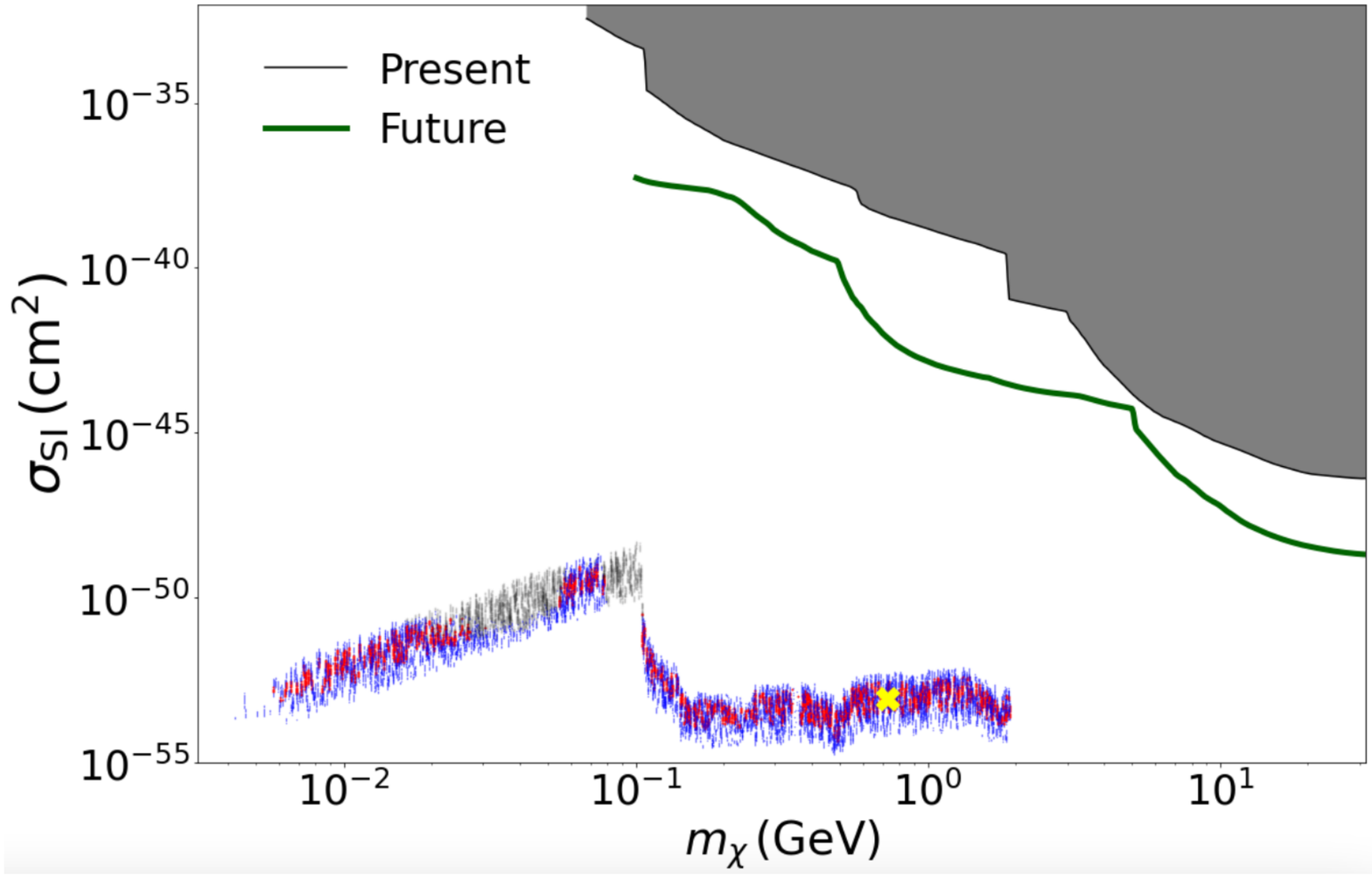}
	\caption{\small \sl The prediction of the scalar dark matter model obtained by casting the favored parameter sets in section\,\ref{subsec: likelihood result} onto the plane of the dark matter mass and the SI scattering cross-section with $C_{h \chi \chi} =$ 0. The colors used for the prediction are the same as those used in Figs.\,\ref{fig: likelihood} and \ref{fig: COL}, except the gray points that indicate the parameter sets excluded by the collider dark matter detection in Fig.\,\ref{fig: COL}. The constraint from present direct dark matter detection experiments and expected sensitivities from future experiments are shown as a gray-shaded region and a dark-green line, respectively.}
	\label{fig: DD}
\end{figure}

Fig.\,\ref{fig: DD} shows the prediction of the scalar dark matter model, where it is obtained by casting the favored parameter sets discussed in the previous section onto the plane of the dark matter mass $m_\chi$ and the SI scattering cross-section $\sigma_{\rm SI}$, with the coupling $C_{h \chi \chi} =$ 0. The colors used for the prediction are the same as those used in Figs.\,\ref{fig: likelihood} and \ref{fig: COL}, except the gray points that indicate the parameter sets excluded by the collider dark matter detection in section\,\ref{subsubsec: Mediator production}. The constraint from present direct dark matter detection experiments and expected sensitivities from future experiments are shown in the same figure as a gray-shaded region and a dark-green line, respectively\,\cite{2104.07634}.\footnote{The most stringent constraint on the SI scattering cross-section at present is obtained from CDEX\,\cite{CDEX:2019hzn} for $m_\chi \lesssim$ 0.1\,GeV, DarkSide-50\,\cite{DarkSide:2018bpj} for 2\,GeV $\lesssim m_\chi \lesssim$ 3\,GeV, and XENON1T\,\cite{XENON:2018voc, XENON:2019zpr, XENON:2020gfr} for other regions, respectively, where the Migdal effect plays an important role in low $m_\chi$ regions\,\cite{Ibe:2017yqa}. On the other hand, the most sensitive projection in the future will be from NEWS-G\,\cite{Durnford:2021mzg} for $m_\chi \lesssim$ 0.5\,GeV, SuperCDMS\,\cite{SuperCDMS:2016wui} for 0.5\,GeV $\lesssim m_\chi \lesssim$ 1.5\,GeV, CYGNUS\,\cite{Vahsen:2020pzb} for 1.5\,GeV $\lesssim m_\chi \lesssim$ 5\,GeV, and DARWIN\,\cite{Schumann:2015cpa} for $m_\chi \gtrsim$ 5\,GeV, respectively.} It is seen in the figure that the strength of the signal is well below the future sensitives, as well as the present constraint.

On the other hand, when the coupling $C_{h \chi \chi}$ becomes non-zero, the direct detection signal is expected to be efficiently enhanced, as its contribution to the scattering cross-section is not suppressed by the small mixing $\sin^2\theta$, as seen in eq.\,(\ref{eq: DD}). However, it turns out that the strength is, at most, comparable to the future sensitivities when $C_{h \chi \chi} \leq {\cal O}(1)$\,GeV.

\subsection{Indirect detection of dark matter}
\label{subsec: ID}

The scalar dark matter is expected to be efficiently searched for at the indirect dark matter detection, as the signal strength is proportional to the annihilation cross-section boosted by the $s$-channel resonance. However, the signal strength also depends on the dark matter density in target materials, which is often a source of the astrophysical uncertainties. Taking those into account, we discuss below the present status and the prospects of scalar dark matter scenario at the indirect detection utilizing cosmic-ray and X/gamma-ray observations.

\subsubsection{Cosmic-ray observations}
\label{subsubsec: CR}

When the scalar dark matter is lighter than ${\cal O}(1)$\,GeV, the contribution to the cosmic-ray flux is from its annihilation into electrons and positrons. However, the produced electrons and positrons have, at most, sub-GeV energies, so those cannot enter the heliosphere due to the solar magnetic field\,\cite{Boudaud:2016mos}. Hence, on the experimental side, only Voyager\,I, i.e., the only cosmic-ray detector located outside the heliosphere since 2012, can detect the cosmic-ray signal. On the other hand, on the theoretical side, the cosmic-ray signal is estimated by considering the injected $e^\pm$ spectra from the dark matter annihilation and its propagation in the Milky Way. The signal flux is proportional to the dark matter density squared and the dark matter annihilation cross-section averaged by the dark matter velocity distribution. Since energetic electrons and positrons quickly lose their energies during the propagation due to their interactions with the interstellar medium, those produced in the vicinity of the solar system mainly contribute to the observed flux. We hence adopt the local dark matter density $\rho_{\rm DM}(r_\odot) = 0.25 \pm 0.11$\,GeV/cm$^3$\,\cite{Read:2014qva}\footnote{The local dark matter density $\rho_{\rm DM}(r_\odot)$ is obtained based on recent local measurements\,\cite{Bovy:2012tw, Garbari:2012ff, Zhang:2012rsb, Bovy:2013raa} without using global measurements assuming spherical symmetry. See the reference \,\cite{Read:2014qva} for more details.} and the local dark matter velocity $v_0(r_\odot) \simeq 300$\,km/s\,\cite{Lacroix:2020lhn}\footnote{Since the effect of the uncertainty from the local dark matter velocity on the signal flux is negligibly small compared to that from the local dark matter density, we do not take it into account in our analysis.} to estimate the signal from the dark matter annihilation.

\begin{figure}[t]
	\centering
    \includegraphics[width=0.6\linewidth]{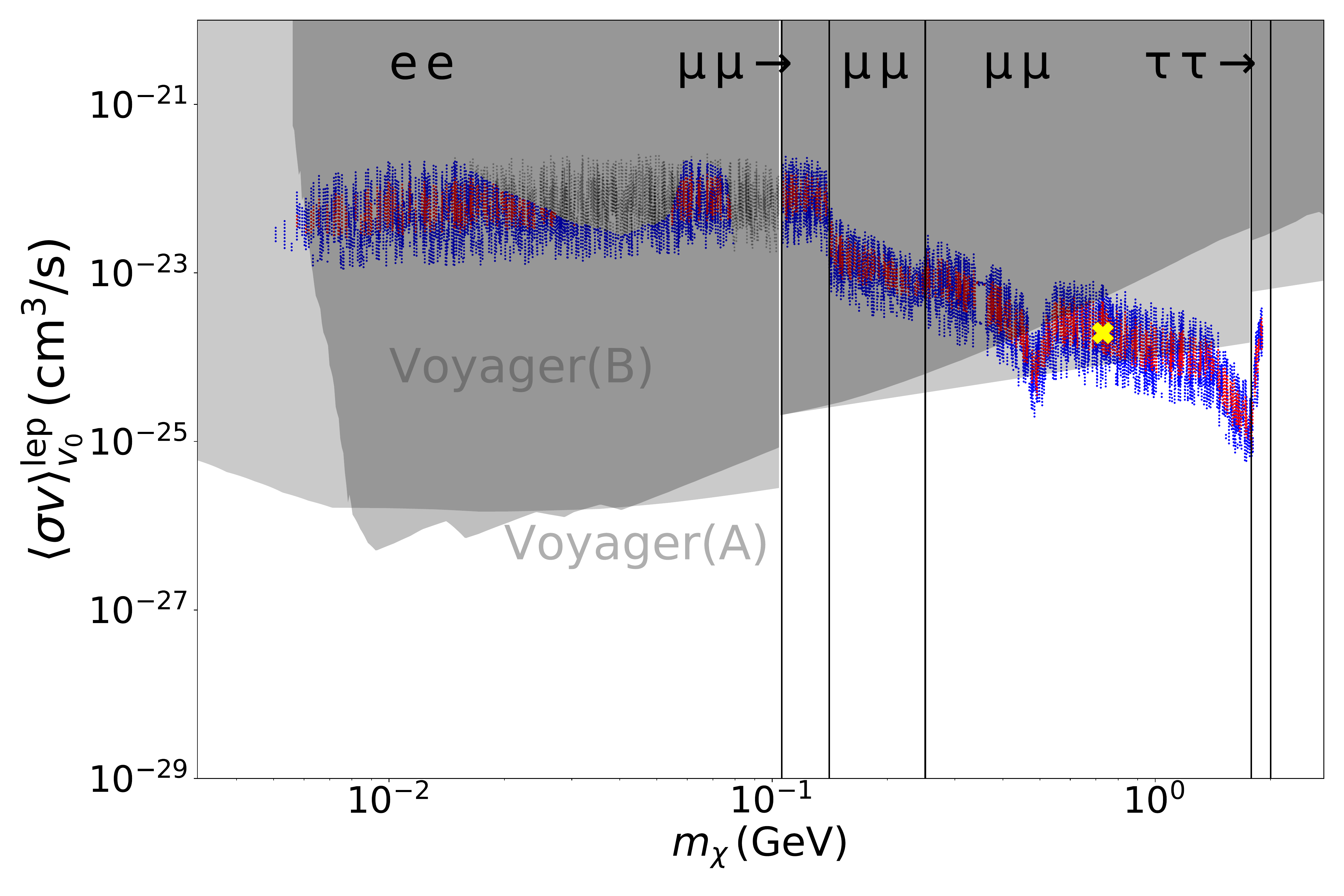}
	\caption{\small \sl The prediction of the scalar dark matter model obtained by casting the favored parameter sets discussed in section\,\ref{subsec: likelihood result} onto the plane of the dark matter mass and the annihilation cross-section into SM lepton pairs. The colors used for the prediction are the same as those used in Fig.\,\ref{fig: DD}. The constraint obtained by the Voyager\,I observation at 90\,\% C.L. is shown by gray-shaded regions, where the "Propagation A" and "Propagation B" models\,\cite{Donato:2003xg, Boudaud:2016mos, Kappl:2015bqa}, which describe the electrons/positrons propagation in our galaxy, are adopted for those in light- and dark-gray shaded regions.}
	\label{fig: cosmic rays}
\end{figure}

Fig.\,\ref{fig: cosmic rays} shows the prediction of the scalar dark matter model obtained by casting the favored parameter sets discussed in Sec\,\ref{subsec: likelihood result} onto the plane of the dark matter mass $m_\chi$ and the annihilation cross-section into SM lepton pairs $\langle \sigma v\rangle_{v_0}^{\rm (lep)} \equiv \sum_\ell \langle \sigma v\,(\chi \chi \to \ell^- \ell^+) \rangle_{v_0}$. The colors used for the prediction are the same as those in Fig.\,\ref{fig: DD}. On the other hand, the constraint from the observation of Voyager\,I at 90\,\% C.L. is shown as gray-shaded regions, where those are obtained using the method developed in Ref.\,\cite{Boudaud:2016mos} assuming that the dark matter annihilates into an electron pair, a muon pair, and a tau lepton pair when the dark matter mass is $m_\chi \leq m_\mu$, $m_\mu \leq m_\chi \leq m_\tau$, and, $m_\chi \geq m_\tau$, respectively. We have considered the uncertainty from the local dark matter density mentioned above and the propagation model to make the constraint conservative and robust. Here, we have adopted the so-called "Propagation A" and "Propagation B" models\,\cite{Donato:2003xg, Boudaud:2016mos, Kappl:2015bqa,Reinert:2017aga}, which describe the propagation of electrons and positrons in our galaxy, to draw the constraints in the figure shown by light- and dark-gray shaded regions, respectively. The "Propagation B" model, which is adopted in Ref.\,\cite{Boudaud:2016mos, Reinert:2017aga} as the model giving the best fit to the latest AMS-02 B/C data, while the “Propagation A” model, which is basically the model MAX of Ref.\,\cite {Donato:2003xg, Maurin:2001sj}, gives the strongest dark matter signal with propagation parameters being consistent with the cosmic-ray data. The main difference between the two models is from the reacceleration; the reacceleration in the "Propagation A" model is stronger, so the constraint using the "Propagation A" is more severe than that using the "Propagation B" model.\footnote{The constraint using the “Propagation A” model is somewhat weaker than that using the “Propagation B” model at $m_\chi = {\cal O}(10)$\,MeV, as the reacceleration up-scatters low energy electrons/positrons to higher ones.} In the following discussion, we have adopted the conservative constraint, namely the one using the “Propagation B” model.

As shown by the letters in the figure, the dark matter annihilates into an electron pair and a muon pair at almost 100\,\% branching fractions when $m_\chi \leq m_\mu$ and $m_\mu \leq m_\chi \leq m_\pi$, respectively. It is found that the Voyager\,I observation does not favor these mass regions. On the other hand, when $m_\pi \leq m_\chi \leq$ 250\,MeV, the main annihilation channel of the dark matter into a SM lepton pair is $\chi \chi \to \mu \mu$, as shown in the figure. However, the dark matter also annihilates into a pair of pions in this region. Though the contribution to the cosmic-ray signal from the latter channel can be estimated using the HAZMA code, we do not include it in our analysis, as this mass region is already ruled out without having such an additional contribution. Finally, the main annihilation channel of the dark matter into a SM lepton pair is $\chi \chi \to \mu \mu$ and $\chi \chi \to \tau \tau$, when 250\,MeV $\leq m_\chi \leq m_\tau$ and $m_\tau \leq m_\chi \leq$ 2\,GeV, respectively, as explicitly mentioned in the figure. The dark matter in these mass regions also annihilates hadronically; however, their contributions to the cosmic-ray signal cannot be robustly estimated because of the same reason addressed in section\,\ref{subsec: CMB}. So, we put the Voyager\,I constraint taking only the leptonic channels into account. It is seen in the figure that several parameter sets survive within the mass region, 250\,MeV $\leq m_\chi \leq$ 2\,GeV.

\subsubsection{MeV gamma-ray observations}
\label{subsubsec: gamma ray}

The scalar dark matter also contributes to the gamma-ray flux through various annihilation channels. However, on the experimental side, the produced photons typically have MeV energies, and this energy range is also known to be the one not intensively explored compared to other energy ranges. The detector’s response is generally no longer point-like for MeV gamma-rays. Moreover, unlike the GeV energy range, the background against the signal in the MeV energy range is more complicated, which includes, e.g., internal, atmospheric, and diffuse gamma-ray components\,\cite{Takada:2021iug}. Despite this difficulty, among various observations, the COMPTEL experiment\,\cite{Strong:1998ck} has performed the most sensitive observation of this energy range, enabling us to explore the MeV gamma-ray signals from the light dark matter annihilation. So, we consider this experiment observing the center of our galaxy to put a robust constraint on the scalar dark matter based on the method developed in Refs.\,\cite{Coogan:2021sjs, Coogan:2021rez}.

On the other hand, on the theoretical side, the gamma-ray signal, i.e., the gamma-ray flux from the dark matter annihilation at the galactic center, is estimated by the formula,
\begin{align}
    \frac{d\Phi_\gamma}{dE_\gamma} \simeq
    \left[
        \frac{\langle \sigma v \rangle_{v_0}}{8 \pi m_\chi^2} \sum_{f_{\rm SM}}\,{\rm Br}\,(\chi \chi \to f_{\rm SM})\,\left.\frac{d N_\gamma}{dE_\gamma}\right|_{f_{\rm SM}}
    \right]
    \times
    \left[
        \int_{\Delta \Omega} d\Omega \int_{\rm l.o.s} ds\,\rho_{\rm DM}^2
    \right],
    \label{flux}
\end{align}
where $\langle \sigma v \rangle_{v_0}$ is the total annihilation cross-section (times the relative velocity) of the dark matter averaged by its relative-velocity distribution with the parameter $v_0 = 400$\,km/s\,\cite{Lacroix:2020lhn}. The branching fraction of the annihilation channel into the SM final state `$f_{\rm SM}$' is denoted by ${\rm Br}\,(\chi \chi \to f_{\rm SM})$, while $d N_\gamma/dE_\gamma|_{f_{\rm SM}}$ is the fragmentation function describing the number of produced photons with energy $E_\gamma$ at a given final state `$f_{\rm SM}$'. The term in the second parenthesis on the right-hand side of the formula is called the J-factor, determined by the dark matter density profile at the galactic center. Among various estimates of the profile\,\cite{Benito:2020lgu, deSalas:2019pee, Cautun:2019eaf}, we adopt the one developed in Ref.\,\cite{Cautun:2019eaf}.\footnote{The so-called NFW dark matter profile was adopted in this reference. So, the J-factor could be estimated more conservatively when we assume a cored dark matter profile rather than the cuspy one. We nevertheless use the estimate using the NFW profile in our analysis, as the constraint from the COMPTEL observation is weaker than that from the Voyager observation even if we adopt it, as seen in the following discussion.} Taking the uncertainty of the J-factor into account, which originates in those of the dark matter profile in the reference, we give a robust constraint on the scalar dark matter from the COMPTEL observation.

\begin{figure}[t]
	\centering
    \includegraphics[width=0.6\linewidth]{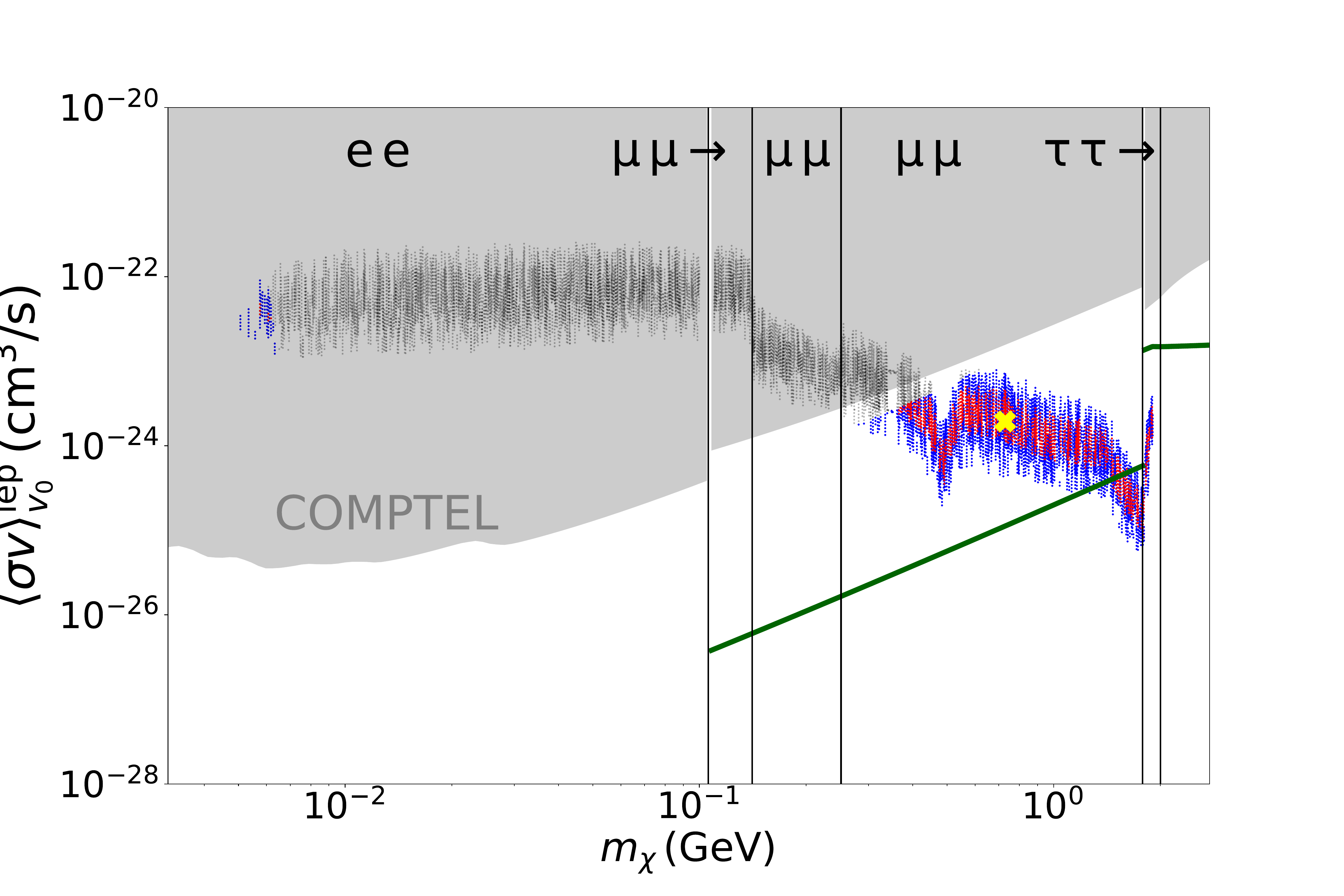}
	\caption{\small \sl The prediction of the scalar dark matter model obtained by casting the favored parameter sets discussed in section\,\ref{subsec: likelihood result} onto the plane of the dark matter mass and the annihilation cross-section into SM lepton pairs. The colors used for the prediction are the same as those used in Fig.\,\ref{fig: cosmic rays}, except that the spread of the gray shaded area is more than in Fig.\,\ref{fig: cosmic rays} because of the Voyager\,I constraint. The 90\,\% C.L. constraint obtained by the COMPTEL observation and the expected sensitivity from the GECCO observation are shown by gray-shaded region and a dark-green line, respectively.}
	\label{fig: gamma ray}
\end{figure}

Fig.\,\ref{fig: gamma ray} shows the prediction of the scalar dark matter model obtained by casting the favored parameter sets discussed in Sec\,\ref{subsec: likelihood result} onto the plane of the dark matter mass $m_\chi$ and the annihilation cross-section into SM lepton pairs $\langle \sigma v\rangle_{v_0}^{\rm (lep)}$. The colors used for the prediction are the same as those in Fig.\,\ref{fig: cosmic rays}, except that the spread of the gray shaded area is more because of the Voyager\,I constraint. The constraint from the COMPTEL observation at 90\,\% C.L. is shown as a gray-shaded region, assuming the same annihilation channels as those in Fig.\,\ref{fig: cosmic rays} in each dark matter mass range. It is seen in the figure that, though the COMPTEL observation is sensitive to exploring the scalar dark matter with the mass below ${\cal O}(100)$\,MeV, its constraint on the dark matter heavier than ${\cal O}(100)$\,MeV is slightly weaker than that from the Voyager observation. On the other hand, future MeV gamma-ray observations\,\cite{Aramaki:2022zpw}, such as the COSI\,\cite{Tomsick:2021wed} and GECCO\,\cite{Orlando:2021get} observations, are expected to efficiently prove this mass range. We also put an expected sensitivity from the GECCO observation in the figure using the method developed in Refs\,\cite{Coogan:2021sjs, Coogan:2021rez}, assuming 3 years observation of the galactic center\,(10\,$\times$\,10 degrees).\footnote{The GECCO sensitivity on the annihilation channel into a tau lepton pair becomes worse than that on the muon pair channel at around $m_\chi \sim m_\tau$. GECCO observes $\gamma$-rays with the energy of $E_\gamma \lesssim$ 8\,MeV, where the main contribution is from the final state radiation associated with a lepton pair production. It is contrary to the case of the COMPTEL constraint, where it is stronger for the annihilation channel into a tau lepton pair than that for the muon pair channel. COMPTEL observes $\gamma$-rays with 1\,MeV $\lesssim E_\gamma \lesssim$ 80\,MeV, where photons from the tau lepton decay mainly contribute to the signal rather than the final state radiation.} It is seen in the figure that the near future observations can almost cover the surviving parameter region in the range of 250\,MeV $\leq m_\chi \leq$ 2\,GeV.

The hard X-ray observation is another exciting possibility for detecting the dark matter, as pointed out in Ref.\,\cite{Cirelli:2020bpc}. Since many energetic electrons and positrons are expected from the annihilation of the scalar dark matter, those produce the X-ray by hitting the interstellar radiation field (CMB, IR, optical photons, etc.) when the dark matter mass is in the sub-GeV. We found that the present constraint from the INTEGRAL X-ray observation on the scalar dark matter is compatible with that from the Voyager observation if we consider several astrophysical ambiguities suggested in the reference. Since a more accurate estimate of the ambiguities is still under debate, it is currently not easy to use the observation to put a robust limit. On the other hand, the hard X-ray observation will be crucial in the future to detecting the sub-GeV dark matter once the ambiguities are well-evaluated because the future MeV gamma-ray observatories will also have a good sensitivity on the hard X-ray range.

\section{Summary}
\label{sec: conclusion}

The scenario with a light scalar dark matter ($\chi$) and a light scalar mediator particle ($\phi$) resolves the so-called core-cusp problem in our universe if the dark matter self-scattering occurs via the Breit-Wigner resonance caused by exchanging the mediator particle in the $s$-channel, namely, $m_\phi \simeq 2 m_\chi$. We have considered the minimal model of the scenario and investigated if it is compatible with the thermal dark matter scenario where the observed dark matter density at the present universe is entirely explained by the freeze-out mechanism of the scalar dark matter. First, we quantitatively figure out the model parameter region (parameter sets) satisfying the thermal relic abundance condition and the self-scattering condition for solving the core-cusp problem, being consistent with the CMB constraints that generally put a very severe limit on light dark matter candidates. We found that such parameter sets reside in the dark matter mass range of 10\,MeV $\leq m_\chi \leq$ 2\,GeV.

Next, we consider the present status and prospects for detecting the light scalar dark matter at the collider, direct and indirect dark matter detections. We immediately found that the signal strength at the direct dark matter detection is too weak to detect, even in the future experiments. On the other hand, we found that some parameter sets have already been ruled out by the collider dark matter detection due to the intensive search for rare $K$-meson decays. Moreover, future $K$-meson experiments can explore most of the parameter sets with $m_\chi \leq$ 100\,MeV. However, it also turns out that a lighter dark matter region, $m_\chi \lesssim$ 300\,MeV, has already been excluded by the indirect dark matter detection using cosmic-ray and gamma-ray observations, for the signal strength is boosted by the $s$-channel resonance mentioned above. It is then found only the parameter sets with 300\,MeV $\lesssim m_\chi \lesssim$ 2\,GeV avoid the severe constraints. This is because, in addition to astrophysical (e.g., the dark matter profile, etc.) and theoretical (e.g., the fragmentation function, etc.) uncertainties, the typical energy of comic-ray and gamma-ray signals at the indirect dark matter detection is in the MeV range, which is known to be the one not intensively explored compared to other ranges. Fortunately, the MeV-energy range will be intensively studied by near future gamma-ray observations, such as COSI\,\cite{Tomsick:2021wed} and GECCO\,\cite{Orlando:2021get} observations, and it enables us to test the scalar dark matter within the mass range of 300\,MeV $\lesssim m_\chi \lesssim$ 2\,GeV.

\appendix

\section{Interactions among Higgs boson and mediator particle}
\label{app: scalar interactions}

\begin{eqnarray}
	C_{h h h} &=& 3\lambda_H v_H c_\theta^3 - 3\mu_{\Phi H} c_\theta^2 s_\theta - \mu_3 s_\theta^3
	+ 3\lambda_{\Phi H} v_H c_\theta s_\theta^2, \nonumber \\
	C_{\phi h h} &=& 3\lambda_H v_H c_\theta^2 s_\theta + \mu_{\Phi H} (c_\theta^3 - 2c_\theta s_\theta^2)
	+ \mu_3 c_\theta s_\theta^2 + \lambda_{\Phi H} v_H (s_\theta^3 - 2c_\theta^2 s_\theta), \nonumber \\
	C_{\phi \phi h} &=& 3\lambda_H v_H c_\theta s_\theta^2 + \mu_{\Phi H} (2c_\theta^2 s_\theta - s_\theta^3)
	- \mu_3 c_\theta^2 s_\theta + \lambda_{\Phi H} v_H (c_\theta^3 - 2c_\theta s_\theta^2), \nonumber \\
	C_{\phi \phi \phi } &=& 3\lambda_H v_H s_\theta^3 + 3\mu_{\Phi H} c_\theta s_\theta^2
	+ \mu_3 c_\theta^3 + 3\lambda_{\Phi H} v_H c_\theta^2 s_\theta, \nonumber \\
	C_{h h h h} &=& 3\lambda_H c_\theta^4 + 6\lambda_{\Phi H} c_\theta^2 s_\theta^2 + \lambda_\Phi s_\theta^4, \nonumber \\
	C_{\phi h h h} &=& 3\lambda_H c_\theta^3 s_\theta - 3\lambda_{\Phi H} (c_\theta^3 s_\theta - c_\theta s_\theta^3)
	- \lambda_\Phi c_\theta s_\theta^3, \nonumber \\
	C_{\phi \phi h h} &=& 3\lambda_H c_\theta^2 s_\theta^2 +\lambda_{\Phi H} (c_\theta^4 - 4c_\theta^2 s_\theta^2 + s_\theta^4)
	+ \lambda_\Phi c_\theta^2 c_\theta^2, \nonumber \\
	C_{\phi \phi \phi h} &=& 3\lambda_H c_\theta s_\theta^3 + 3\lambda_{\Phi H} (c_\theta^3 s_\theta - c_\theta s_\theta^3)
	- \lambda_\Phi c_\theta^3 s_\theta, \nonumber \\
	C_{\phi \phi \phi \phi} &=& 3\lambda_H s_\theta^4 + 6\lambda_{\Phi H} c_\theta^2 s_\theta^2 + \lambda_\Phi c_\theta^4,
\end{eqnarray}

\section*{Acknowledgments}

This work is supported by Grant-in-Aid for Scientific Research from the Ministry of Education, Culture, Sports, Science, and Technology (MEXT), Japan; 20H01895, 20H00153, 19H05810,18H05542 (for S. Matsumoto), by World Premier International Research Center Initiative (WPI), MEXT, Japan (Kavli IPMU), by the JST SPRING, Grant Number JPMJSP2108, and by JSPS Core-to-Core Program (JPJSCCA20200002). SC also acknowledges support from the Indo-French Centre for the Promotion of Advanced Research (CEFIPRA Grant no: 6304-2).


\bibliographystyle{unsrt}
\bibliography{refs}

\end{document}